\documentclass[12pt,a4paper]{article}
\usepackage{natbib,graphicx,amssymb,bm,setspace,dcolumn,amsmath,todonotes}

\renewcommand\footnotemark{}

\title{Cross-validatory extreme value threshold selection and uncertainty with application to ocean storm severity
}
\author{Paul J. Northrop 
\footnote{
{\it Address for correspondence}: Paul Northrop, Department of Statistical Science, University College London, Gower Street, London WC1E 6BT, UK. 
E-mail: p.northrop@ucl.ac.uk}
and Nicolas Attalides \\
{\it University College London, UK} \\ \\
Philip Jonathan \\
{\it Shell Projects \& Technology, Manchester, UK} 
}

\let\leq=\leqslant
\let\geq=\geqslant
\let\hat=\widehat
\newcommand{\bit}{\begin{itemize}}
\newcommand{\eit}{\end{itemize}}
\newcommand{\beqnn}{\begin{eqnarray*}}
\newcommand{\eeqnn}{\end{eqnarray*}}
\newcommand{\beqn}{\begin{eqnarray}}
\newcommand{\eeqn}{\end{eqnarray}}
\newcommand{\bx}{\bm{x}}
\newcommand{\simiid}{\,\stackrel{{\rm\tiny i.i.d.}}{\sim}\,}

\def\dperp{
\mathrel{%
\kern0pt\vbox{\hbox to 0.8em{\hss
\setbox0=\hbox{$\shortparallel$}\dp0=0pt\box0\hss}\hrule width 0.8em}%
}}

\begin{document}
\maketitle
\begin{abstract}
Designs conditions for marine structures are typically informed by threshold-based extreme value analyses of oceanographic variables, in which
excesses of a high threshold are modelled by a generalized Pareto (GP) distribution.  
Too low a threshold leads to bias from model mis-specification; raising the threshold increases the variance of estimators: a bias-variance trade-off. 
Many existing threshold selection methods do not address this trade-off directly, but rather aim to select the lowest threshold above which the GP model is judged to hold approximately.
In this paper Bayesian cross-validation is used to address the trade-off by comparing thresholds based on predictive ability at extreme levels.
Extremal inferences can be sensitive to the choice of a single threshold.
We use Bayesian model-averaging to combine inferences from many thresholds, thereby reducing sensitivity to the choice of a single threshold.
The methodology is applied to significant wave height datasets from the northern North Sea and the Gulf of Mexico.
\end{abstract}


\section{Introduction}
Ocean and coastal structures, including breakwaters, ships and oil and gas producing facilities are designed to withstand extreme environmental conditions. Marine engineering design codes stipulate that estimated failure probabilities of offshore structures, associated with one or more return periods, should not exceed specified values. To characterize the environmental loading on an offshore structure, return values for winds, waves and ocean currents corresponding to a return period of typically 100 years, but sometimes to 1,000 and 10,000 years are required.  The severity of waves in a storm is quantified using significant wave height.  Extreme value analyses of measured and hindcast samples of significant wave height are undertaken to derive environmental design criteria, typically by fitting a GP distribution to excesses of a high threshold.  The selection of appropriate threshold(s) is important because inferences can be sensitive to threshold.     

\subsection{Storm peak significant wave height datasets}
\begin{figure}[h]
\centering
\includegraphics[width=0.9\textwidth, angle=0]{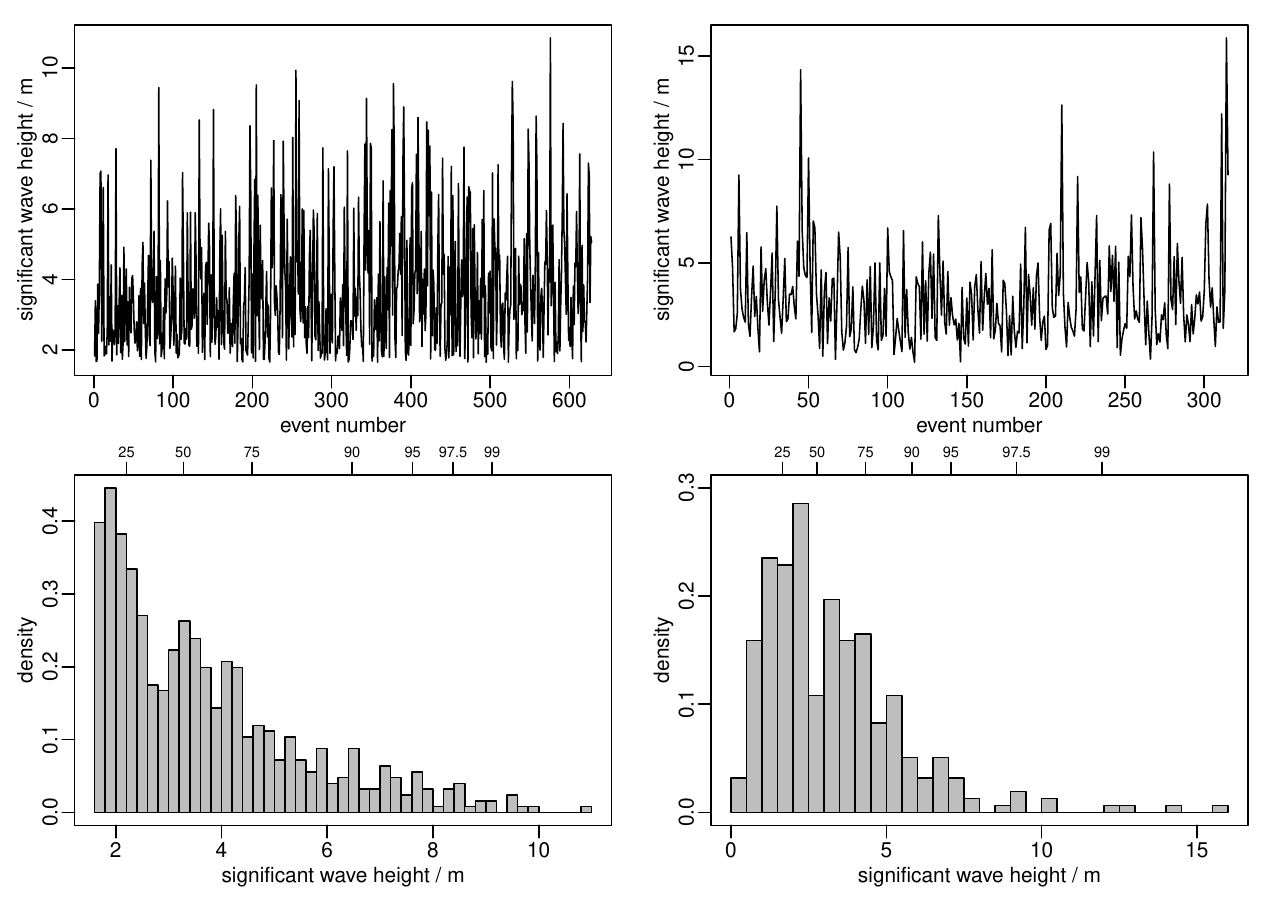}
\vspace{-0.25cm}
\caption{\label{fig:Hs_data} Storm peak significant wave height hindcast datasets.  
Left: North Sea data (628 observations).  Right: Gulf of Mexico data (315 observations).
Top: time series plots.  Bottom: histograms. The upper axis scales give the sample quantile levels.}
\end{figure}

The focus of this paper is the analysis of two sequences of hindcasts of storm peak significant wave height, shown in Figure \ref{fig:Hs_data}.  
Significant wave height ($H_s$) is a measure of sea surface roughness.
It is defined as four times the standard deviation of the surface elevation of a {\it sea state}, the ocean surface observed for a certain period of time (3 hours for our datasets).  \cite{CCCCMS2014} gives the largest $H_s$ value ever generated by a hindcast model as 18.33m and the largest value ever measured in the ocean as 20.63m.
Hindcasts are samples from physical models of the ocean environment, calibrated to observations of pressure, wind and wave fields.

For each of the datasets raw data have been declustered, using a procedure described in \cite{EJ2008}, to isolate cluster maxima (storm peaks) that can reasonably be treated as being mutually independent.  We also assume that storm peaks are sampled from a common distribution. 
Even in this simplest of situations practitioners have difficulty in selecting appropriate thresholds.
The first dataset, from an unnamed location in the northern North Sea, contains 628 storm peaks from October 1964 to March 1995, but restricted to the period October to March within each year.
The other dataset contains 315 storm peaks from September 1900 to September 2005.  

In the northern North Sea the main fetches are the Norwegian Sea to the North, the Atlantic Ocean to the west, and the North Sea to the south. 
Extreme sea states from the directions of Scandinavia to the east and the British Isles to the southwest are not possible, owing to the shielding effects of these land masses.
At the location under consideration, the most extreme sea states are associated with storms from the Atlantic Ocean \citep{EJ2008}.  With up to several tens of storms impacting the North Sea each winter, the number of events for analysis is typically larger than for locations in regions such as the Gulf of Mexico, where hurricanes produce the most severe sea states. 
Most hurricanes originate in the Atlantic Ocean between June and November and propagate west to northwest into the Gulf producing the largest sea states with dominant directions from the southeast to east directions. 
It is expected that there is greater potential for very stormy sea conditions in the Gulf of Mexico than in the northern North Sea and therefore that the extremal behaviour is different in these two locations.

\subsection{Extreme value threshold selection}
Extreme value theory provides asymptotic justification for a particular family of models for excesses of a high threshold.
Let $X_1, X_2, \ldots X_n$ be a sequence of independent and identically distributed random variables, with common distribution function $H$, and $u_n$ be a threshold, increasing with $n$.  
\cite{Pickands1975} showed that if there is a non-degenerate limiting
distribution for appropriately linearly rescaled excesses of $u_n$ then this limit is a Generalized Pareto (GP) distribution.
In practice, a suitably high threshold $u$ is chosen empirically.
Given that there is an exceedance of $u$, the excess $Y=X-u$ is modelled by a GP($\sigma_u,\xi$) distribution, with positive threshold-dependent scale parameter $\sigma_u$, shape parameter $\xi$ and distribution function
\begin{equation}
G(y)= 
\begin{cases} 
1-\left( 1+\xi y /\sigma_u\right)_+^{-1/\xi}, & \xi \neq 0, \\
1-\exp(-y/\sigma_u), &  \xi = 0,
\end{cases} \label{eqn:GP}
\end{equation}
where $y>0$, $x_+\!=\!\max(x,0)$.
The $\xi=0$ case is defined in the limit as $\xi \rightarrow 0$.
When $\xi <0$ the distribution of $X$ has a finite upper endpoint of $u-\sigma_u/\xi$; 
otherwise, $X$ is unbounded above.
The frequency with which the threshold is exceeded also matters.
Under the current assumptions the number of exceedances $u$ has a binomial$(n,p_u)$ distribution, where $p_u=P(X>u)$, giving a BGP($p_u,\sigma_u,\xi$) model \citep[chapter 4]{Coles2001}

Many threshold diagnostic procedures have been proposed: \cite{SM2012} provides a review.
Broad categories of methods include:  
assessing stability of model parameter estimates with threshold \citep{DdHR2000,Wadsworth2015};
goodness-of-fit tests \citep{DS90,Dupuis1998}; 
approaches that minimize the asymptotic mean-squared error of estimators of $\xi$ or of extreme quantiles,  
under particular assumptions about the form of the upper tail of $H$ \citep{HW1985, Hall1990, FdHP2003, BGST2004};
specifying a model for (some or all) data below the threshold \citep{WL2010, MSLDRR2011, WT2012}.
In the latter category, the threshold above which the GP model is assumed to hold is treated as a model parameter and {\it threshold uncertainty} is incorporated by averaging inferences over a posterior distribution of model parameters.
In contrast, in a {\it single threshold approach} threshold level is viewed as a tuning parameter, whose value is selected prior to the main analysis and is treated as fixed and known when subsequent inferences are made.

Single threshold selection involves a bias-variance trade-off \citep{SM2012}: the lower the threshold the greater the estimation bias due to model misspecification; the higher the threshold the greater the estimation uncertainty.  
Many existing approaches do not address the trade-off directly, but rather examine sensitivity of inferences to threshold and/or aim to select the lowest threshold above which the GP model appears to hold approximately.
We seek to deal with the bias-variance trade-off based on the main purpose of the modelling, i.e. prediction of future extremal behaviour.  We make use of a data-driven method commonly used for such purposes: cross-validation \citep{Stone1974}.
We consider only the simplest of modelling situations, i.e. where observations are treated as independent and identicaly distributed.
However, selecting the threshold level is a fundamental issue for all threshold-based extreme value analyses and we anticipate that our general approach can have much wider applicability.

\begin{figure}[h]
\centering
\includegraphics[width=0.9\textwidth, angle=0]{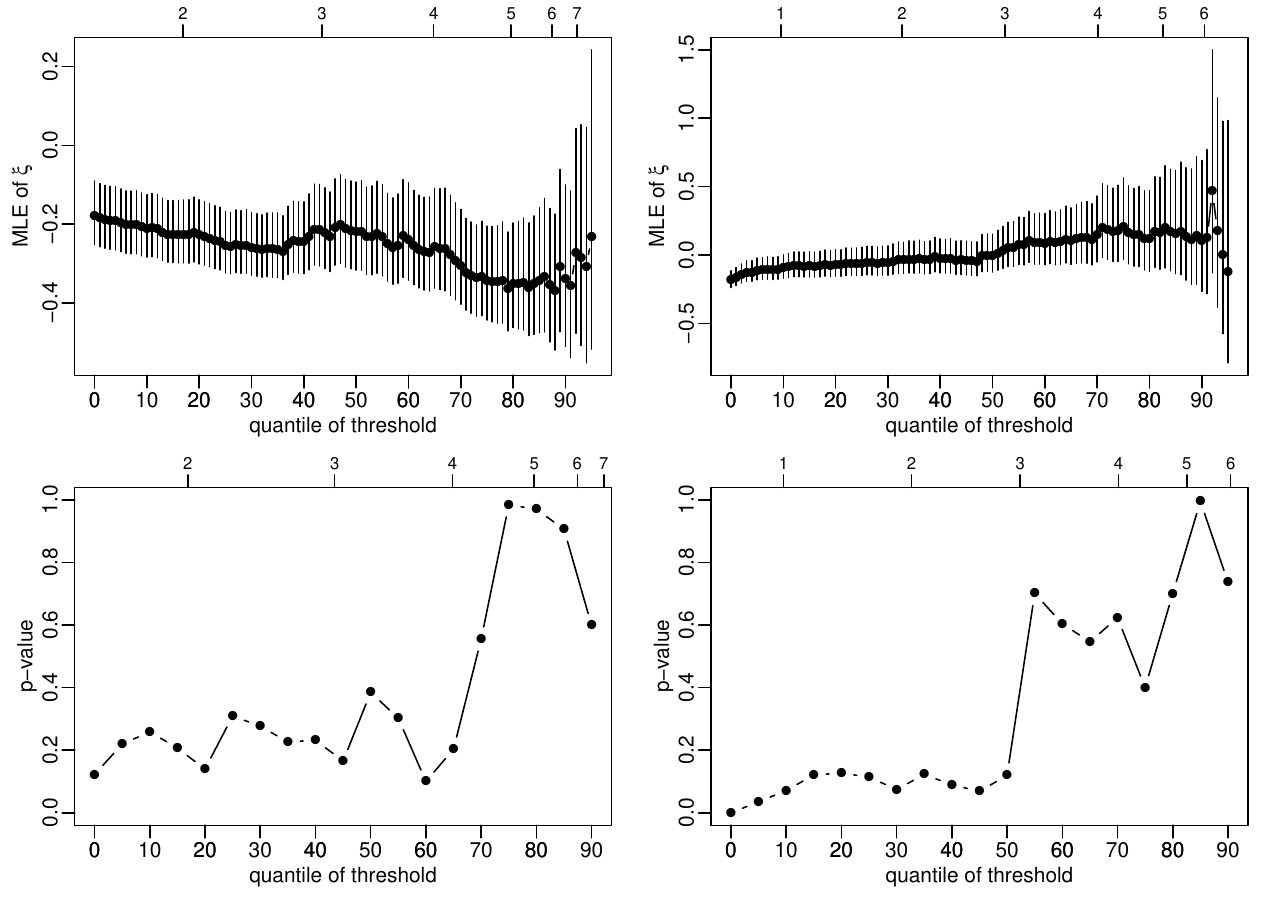}
\vspace{-0.25cm}
\caption{\label{fig:thresh_diag}  Threshold diagnostic plots for the storm peak significant wave height hindcast datasets.  
Left: North Sea data.  Right: Gulf of Mexico data.
Top: parameter stability plots for MLEs of $\xi$, with 95\% pointwise profile likelihood-based confidence intervals.
Bottom: $p$-values associated with a test of constant shape parameter against the lowest threshold considered.  
The upper axis scales give the level of the threshold in metres.}
\end{figure}

We illustrate some of the issues involved in threshold selection by applying to the significant wave height datasets two approaches that assess parameter stability.  
In the top row of Figure \ref{fig:thresh_diag} maximum likelihood (ML) estimates $\hat{\xi}$ of $\xi$ are plotted against threshold.  
The aim is to choose the lowest threshold above which $\hat{\xi}$ is approximately constant in threshold, taking into account sampling variability summarized by the confidence intervals.  It is not possible to make a definitive choice and different viewers may choose rather different thresholds.  
In both of these plots {\it our} eyes are drawn to the approximate stability of the estimates at around the 70\% sample quantile.
One could argue for lower thresholds, to incur some bias in return for reduced variance, but it is not possible to assess this objectively from these plots.
In practice, it is common not to consider thresholds below the apparent mode of the data because the GP distribution has it's mode at the origin.
For example, based on the histogram of the Gulf of Mexico data in Figure \ref{fig:Hs_data} one might judge a threshold below the 25\% sample quantile to be unrealistic.
However, we {\it will} consider such thresholds because it is interesting to see to what extent the bias expected is offset by a gain in precision.

The inherent subjectivity of this approach, and other issues such as the strong dependence between estimates of $\xi$ based on different thresholds, motivated more formal assessments of parameter stability
\citep{WT2012, NC2014, Wadsworth2015}. 
The plots in the bottom row of Figure \ref{fig:thresh_diag} are based on \cite{NC2014}.
A subasymptotic piecewise constant model \citep{WT2012} is used in which the value of $\xi$ may change at each of a set of thresholds, here set at the 0\%, 5\%, \ldots, 95\% sample quantiles. 
For a given threshold the null hypothesis that the shape parameter is constant from this threshold upwards is tested.
In the plots $p$-values from this test are plotted against threshold.
Although these plots address many of the inadequacies of the parameter estimate stability plots subjectivity remains because one must decide how to make use of the $p$-values.  
One could prespecify a size, e.g. 0.05, for the tests or take a more informal approach by looking for a sharp increase in $p$-value.
For the North Sea data the former would suggest a very low threshold and the latter a threshold in the region of the 70\% sample quantile.
For the Gulf of Mexico data respective thresholds close to the 10\% and 55\% sample quantiles are indicated.

An argument against selecting a single threshold is that this ignores uncertainty concerning the choice of this threshold.
As mentioned above, one way to account for this uncertainty is to embed a threshold parameter within a model.
We use an approach based on Bayesian Model Averaging (BMA), on which \cite{HMRV1999} provide a review.
\cite{SNF2013} have recently used a similar approach to combine inferences from different multivariate extreme value models.
We treat different thresholds as providing competing models for the data.  Predictions of extremal behaviour are averaged over these models, with individual models weighted in proportion to the extent to which they are supported by the data.  There is empirical and theoretical evidence \citep{HMRV1999} that averaging inferences in this way results in better average predictive ability than provided by any single model.

For the most part we work in a Bayesian framework because prediction is handled naturally and regularity conditions required for making inferences using ML \citep{Smith1985} and probability-weighted moments (PWM) \citep{Hosking1987}, namely $\xi>-1/2$ and $\xi<1/2$ respectively, can be relaxed.
This requires a prior distribution to be specified for the parameters of the BGP model.  Initially, we consider three different `reference' prior distributions, 
in the general sense of priors constructed using formal rules \citep{Kass1996}.  
Such priors can be useful when information provided by the data is much stronger than prior information from other sources. 
This is more likely to be the case for a low threshold than for a high threshold. 
We use simulation to assess the utility of these priors for our purpose, that is, making predictive statements about future extreme observations
and use the results to formulate an improved prior.
For high thresholds, when the data are likely to provide only weak information, it may be important to incorporate at least some basic prior information in order to avoid making physically unrealistic inferences.      

In section \ref{sec:CV} we use Bayesian cross-validation to estimate a measure of threshold performance for use in the selection of a single threshold.
In section \ref{sec:BMA} we consider how to use this measure to combine inferences over many thresholds.
Sections \ref{sec:CVthresh} and \ref{comp_thresh} describe the cross-validation procedure and its role in selecting a single threshold.
In section \ref{sec:extreme_pred} we discuss two related formulations of the objective of an extreme value analysis and, in section 
\ref{sec:sim1}, we use one of these formulations in a simulation study to inform the choice of prior distribution for GP parameters.
Another simulation study, in section \ref{sec:sim2}, compares the strategies of choosing a single `best' threshold and averaging of many thresholds.
In sections \ref{sec:real1} and \ref{sec:real2} we use our methodology to make inferences about extreme significant wave heights in the North Sea and in the Gulf of Mexico.  
In section \ref{sec:WIprior} we illustrate how one could incorporate prior information about the GP shape parameter $\xi$ to avoid physically unrealistic inferences.

\section{Single threshold selection}
\label{sec:CV}
We use a Bayesian implementation of leave-one-out cross-validation to compare the predictive ability of BGP inferences based on different thresholds.
We take a predictive approach, averaging inferences over the posterior distribution of parameters,
to reflect differing parameter uncertainties across thresholds: uncertainty in GP parameters will tend to increase as the threshold is raised.  
A point estimate of GP model parameters can give a zero likelihood for a validation observation:
this occurs if $\hat{\xi}<0$ and this observation is greater than the estimated upper endpoint $u-\hat{\sigma}_u/\hat{\xi}$.
In this event an estimative approach would effectively rule out the threshold $u$.  
Accounting for parameter uncertainty alleviates this problem by giving weight to parameter values other than a particular point estimate.

A naive implementation of leave-one-out cross-validation is computationally intensive.  
To avoid excessive computation we use importance sampling to estimate cross-validation predictive densities based on Bayesian inferences from the entire dataset.  
One could use a similar strategy in a frequentist approximation to predictive inference based on large sample theory or bootstrapping \citep[chapter 10]{YS2005}.
However, large sample results may provide poor approximations for high thresholds (small number of excesses) and the GP observed
information is known to have poor finite-sample properties \citep{SD2010}.  Bootstrapping, of ML or PWM estimates, increases computation time further and is subject to the regularity conditions mentioned in the introduction.

\subsection{Assessing threshold performance using cross-validation}
\label{sec:CVthresh}
Suppose that $\bx=(x_1, \ldots, x_n)$ is a random sample of raw (unthresholded) data from $H$.
Without loss of generality we assume that $x_1 < \cdots < x_n$.
Consider a {\it training threshold} $u$.  
A BGP($p_u,\sigma_u,\xi$) model is used at threshold $u$,
where $p_u=P(X>u)$ and ($\sigma_u,\xi$) are the parameters of the GP model for excesses of $u$.  
Let $\theta=(p_u,\sigma_u,\xi)$ and $\pi(\theta)$ be a prior density for $\theta$.
Let $\bx^s$ denote a subset of $\bx$, possibly equal to $\bx$.
The posterior density $\pi_u(\theta~|~\bx^s) \propto L(\theta;\bx^s, u) \pi(\theta)$, where
\[ L(\theta;\bx^s, u)=\prod_{i: x_i \in \bx^s}  f_u(x_i~|~\theta), \]
\[ f_u(x_i~|~\theta) = (1-p_u)^{I(x_i \leq u)} \left\{ p_u g(x_i-u;\sigma_u,\xi) \right\}^{I(x_i > u)}, \]
$I(x)=1$ if $x$ is true and $I(x)=0$ otherwise, and
\[ g(x;\sigma_u,\xi) = \sigma_u^{-1} \left( 1+\frac{\xi x}{\sigma_u} \right)_{\!\!+}^{-(1+1/\xi)}, \]
is the density of a GP$(\sigma_u,\xi)$ distribution.

We quantify the ability of BGP inferences based on threshold $u$ to predict (out-of-sample) at extreme levels.
For this purpose we introduce a {\it validation threshold} $v \geq u$.
If $1+\xi(v-u)/\sigma_u>0$ then a BGP($p_u,\sigma_u,\xi$) model at threshold $u$ implies a BGP($p_v,\sigma_v,\xi$) model at threshold $v$, where
$\sigma_v=\sigma_u+\xi(v-u)$ and
$p_v= P(X>v) = \left( 1+\xi(v-u)/\sigma_u \right)^{-1/\xi} p_u$.
Otherwise, $p_v=0$ and excesses of $v$ are impossible.
For a particular value of $v$ we wish to compare the predictive ability of the implied BGP($p_v,\sigma_v,\xi$) model across a range of values of $u$.

We employ a leave-one-out cross-validation scheme in which $\bx_{(r)}=\{x_i, i\neq r\}$ forms the training data and $x_r$ the validation data.
The {\it cross-validation predictive densities} at validation threshold $v$, based on a training threshold $u$, are given by
\beqn
f_v(x_r~|~\bx_{(r)}, u)&=& \int f_v(x_r~|~\theta, \bx_{(r)}) \,\pi_u(\theta~|~\bx_{(r)}) {\rm ~d}\theta, \quad r=1, \ldots, n. \label{full_CV}
\eeqn
Suppose that the $\{ x_i \}$ are conditionally independent given $\theta$.
If $p_v>0$ then 
\beqn
f_v(x_r~|~\theta, \bx_{(r)}) = f_v(x_r~|~\theta) = (1-p_v)^{I(x_r \leq v)} \left\{ p_v g(x_r-v;\sigma_v,\xi) \right\}^{I(x_r > v)}. \label{eqn:vallik}
\eeqn
If $p_v=0$ then $f_v(x_r~|~\theta, \bx_{(r)}) = I(x_r \leq v)$. 
Suppose that we have a sample $\theta^{(r)}_j, j=1, \ldots, m$ from the posterior $\pi_u(\theta~|~\bx_{(r)})$.  
Then a Monte Carlo estimator of $f_v(x_r~|~\bx_{(r)}, u)$ based on (\ref{full_CV}) is given by
\beqn
\hat{f}_v(x_r~|~\bx_{(r)}, u) = \frac1m \sum_{j=1}^m f_v(x_r~|~\theta^{(r)}_j, \bx_{(r)}).  \label{full_CV_est}
\eeqn
Evaluation of estimator (\ref{full_CV_est}), for $r=1, \ldots, n$, is computationally intensive because it involves generating samples from $n$ different posterior distributions.
To reduce computation time we use an importance sampling estimator \citep{Gelfand1996,GD1994} that enables estimation of 
$\hat{f}_v(x_r~|~\bx_{(r)}, u)$, for $r=1, \ldots, n-1$, using a single sample only. 
We rewrite (\ref{full_CV}) as
\beqn
f_v(x_r~|~\bx_{(r)}, u)&=& \int f_v(x_r~|~\theta, \bx_{(r)}) \, q_r(\theta) \, h(\theta) {\rm ~d}\theta, \quad r=1, \ldots, n, \label{IS_CV}
\eeqn
where $q_r(\theta)=\pi_u(\theta~|~\bx_{(r)})/h(\theta)$ and $h(\theta)$ is a density whose support must include that of $\pi_u(\theta~|~\bx_{(r)})$.
In the current context a common choice is $\pi_u(\theta~|~\bx)$ \citep[page 511]{GD1994}.
However, the support of $\pi_u(\theta~|~\bx)$: $\xi>-\sigma_u/(x_n-u)$,
does not contain that of $\pi_u(\theta~|~\bx_{(n)})$: $\xi>-\sigma_u/(x_{n-1}-u)$, since $x_n > x_{n-1}$.
Therefore we use $h(\theta)=\pi_u(\theta~|~\bx)$ only for $r \neq n$.

Suppose that we have a sample $\theta_j, j=1, \ldots, m$ from the posterior $\pi_u(\theta~|~\bx)$.  For $r=1, \ldots, n-1$ we use the importance sampling ratio estimator
\beqn
\hat{f}_v(x_r~|~\bx_{(r)}, u)
&=& \frac{\sum_{j=1}^m f_v(x_r~|~\theta_j) \, q_r(\theta_j)}
{\sum_{j=1}^m q_r(\theta_j)}, \label{eqn:imp} \\
&=&\frac{
\sum_{j=1}^m f_v(x_r~|~\theta_j) / f_u(x_r~|~\theta_j)}
{\sum_{j=1}^m 1 / f_u(x_r~|~\theta_j) },
\label{quick_CV_est}
\eeqn
where $q_r(\theta)=\pi_u(\theta~|~\bx_{(r)})/\pi_u(\theta~|~\bx) 
\propto 1/f_u(x_r~|~\theta)$.
If we also have a sample $\theta^{(n)}_j, j=1, \ldots, m$ from the posterior $\pi_u(\theta~|~\bx_{(n)})$ then 
$\hat{f}_v(x_n~|~\bx_{(n)}, u)=(1/m)\sum_{j=1}^m f_v(x_n~|~\theta_j^{(n)})$. 

We use 
\begin{equation}
\hat{T}_v(u)=\sum_{r=1}^n \log \hat{f}_v(x_r~|~\bx_{(r)},u) \label{eqn:tw}
\end{equation} as a measure of predictive performance at validation threshold $v$ when using training threshold $u$.

\subsection{Comparing training thresholds}
\label{comp_thresh}
Consider $k$ training thresholds $u_1 < \cdots < u_k$, resulting in estimates $\hat{T}_v(u_1), \ldots, \hat{T}_v(u_k)$.
Up to an additive constant, $\hat{T}_v(u)$ provides an estimate of the negated Kullback-Leibler divergence between the BGP model at validation threshold $v$ and the true density (see, for example, \citet[page 53]{Silverman1986}).
Thus, $u^*=\operatorname{arg\,max}_u \hat{T}_v(u)$ has the property that, of the thresholds considered, it has the smallest estimated Kullback-Leibler divergence.

Some inputs are required: $u_1, \ldots, u_k$, $v$ and $\pi(\theta)$. 
Choosing a set $u_1, \ldots, u_k$ of plausible thresholds that span the range over which the bias-variance trade-off is occurring is the starting point for many threshold selection methods.
An initial graphical diagnostic, such as a parameter stability plot, can assist this choice.  
In particular, $u_k$ should not be so high that little information is provided about GP parameters. 
There is no definitive rule for limiting $u_k$ but \citet{JE2013} suggest that there should be no fewer than 50 threshold excesses. 
Applying this rule would restrict $u_k$ to be no higher than the 84\% and 92\% sample quantiles for the Gulf of Mexico and North Sea datasets respectively, but later we will use $u_k$ that break the rule and examine the consequences.

We need $v \geq u_k$, but the larger $v$ is the fewer excesses of $v$ there are and the smaller the information from data thresholded at $v$.  
Consider two validation thresholds: $v_1=u_k$ and $v_2>u_k$.
If we use $v_2$ we lose validation information: if $v_1 < x_r \leq v_2$ then in (\ref{eqn:vallik}) $x_r$ is censored rather than entering into the GP part of the predictive density; and gain nothing: the prediction of $x_r>v_2$ is unaffected by the choice of $v_1$ or $v_2$ because
$p_{v_1}\,g(x-v_1;\sigma_{v_1};\xi)=p_{v_2}\,g(x-v_2;\sigma_{v_2};\xi)$. 
Therefore, we should use $v=u_k$.

In section \ref{sec:sim1} we compare predictive properties of three `reference' priors for GP parameters.
Such priors are intended for use when substantial prior information is not available and it is anticipated that information provided by the data will dominate the posterior distribution \citep{ohagan2006}.  
For high thresholds this may not be the case, with the possible consequence that a diffuse posterior places non-negligible posterior probability on unrealistically large values of $\xi$ and produces unrealistic extrapolation to long future time horizons.
In this event one needs to incorporate more information, perhaps by using a more considered prior distribution, and/or limit the time horizon of interest.   
We return to this issue in section \ref{sec:WIprior}.

\subsection{Prediction of extreme observations}
\label{sec:extreme_pred}
In an extreme value analysis the main focus is often the estimation of extreme quantiles called {\it return levels}.
Let $M_N$ denote the largest value observed over a time horizon of $N$ years.
The $N$-year return level $z(N)$ is defined as the value exceeded by an annual maximum $M_1$ with probability $1/N$.  
In off-shore engineering design criteria are usually expressed in terms of return levels, for values of $N$ such as 100, 1000, 10,000.
A related approach defines the quantity of interest as the random variable $M_N$, rather than particular quantiles of $M_1$.
Under a BGP($p_u,\sigma_u,\xi$) model, for $z>u$,
\[ F(z;\theta) = P(X \leq z) = 1-p_u \left\{ 1+\xi\left(\frac{z-u}{\sigma_u}\right)\right\}^{-1/\xi}. \]
Then $z(N)=z(N;\theta)$ satisfies $F(z(N);\theta)^{n_y}=1-1/N$, where $n_y$ is the mean number of observations per year.
Similarly, for $z>u$, $P(M_N \leq z)=F(z;\theta)^{n_y N}$.
For large  $N$ ($N=100$ is sufficient), $z(N)$ is approximately equal to the 37\% quantile of the distribution of $M_{N}$ \citep{CIN2002}.
In an estimative approach, based on a point estimate of $\theta$, the value of $z(N)$ is below the median of $M_N$.
A common interpretation of $z(N)$ is the level exceeded on average once every $N$ years.
However, for large $N$ (again $N=100$ is sufficient) and under an assumption of independence at extreme levels, 
$z(N)$ is exceeded 0, 1, 2, 3, 4 times with respective approximate probabilities of 37\%, 37\%, 18\%, 6\% and 1.5\%.
It may be more instructive to examine directly the distribution of $M_N$,
rather than very extreme quantiles of the annual maximum $M_1$.

The relationship between these two approaches is less clear under a predictive approach, in which posterior uncertainty about $\theta$ in incorporated into the calculations.
The $N$-year {\it (posterior) predictive return level} $z_P(N)$ is the solution of
\[ P(M_1 \leq z_P(N)~|~\bx) = \int F(z_P(N);\theta)^{n_y} \, \pi(\theta~|~\bx) {\rm ~d}\theta = 1-1/N, \]
and the predictive distribution function of $M_N$ is given by 
\begin{equation}
P(M_N \leq z~|~\bx) = \int F(z;\theta)^{n_yN} \, \pi(\theta~|~\bx) {\rm ~d}\theta. 
\label{eqn:MNpred}
\end{equation}
As noted by \citet[section 1.3]{Smith2003}, accounting for parameter uncertainty tends to lead to larger estimated probabilities of extreme events, that is, $z_P(N)$ tends to be greater than an estimate $\hat{z}(N)$ based on, for example, the MLE $\hat{\theta}$.
The strong non-linearity of $F(z;\theta)^{n_y}$ for large $z$, and the fact that it is bound above by 1, mean that averages of $F(z;\theta)^{n_y}$ over areas of the parameter space relating to the extreme upper tail of $M_1$ tend to be smaller than point values near the centre of such areas.  This phenomenon is less critical when working with the distribution of $M_N$ because now central quantiles of $M_N$ also have relevance, not just particular extreme tail probabilities.  
Numerical results in section \ref{sec:real1} (Figure \ref{fig:weights}) show that $z_P(N)$ can be rather greater than the median of the predictive distribution of $M_N$, particularly when posterior uncertainty about $\theta$ is large.

For a given value of $N$, we estimate $P(M_N \leq z~|~\bx)$ using the sample $\theta_j, j=1,\ldots, m$ from the posterior density $\pi(\theta~|~\bx)$ to give
\begin{equation}
\hat{P}(M_N \leq z~|~\bx) = \frac1m \sum_{j=1}^m F(z;\theta_j)^{n_yN}.  \label{eqn:MNpredest}
\end{equation}
The solution $\hat{z}_P(N)$ of $\hat{P}(M_1 \leq \hat{z}_P(N)~|~\bx)=1-1/N$ provides an estimate of $z_P(N)$.

\subsection{Simulation study 1: priors for GP parameters}
\label{sec:sim1}
\begin{figure}[h]
\centering
\includegraphics[width=0.495\textwidth, angle=0]{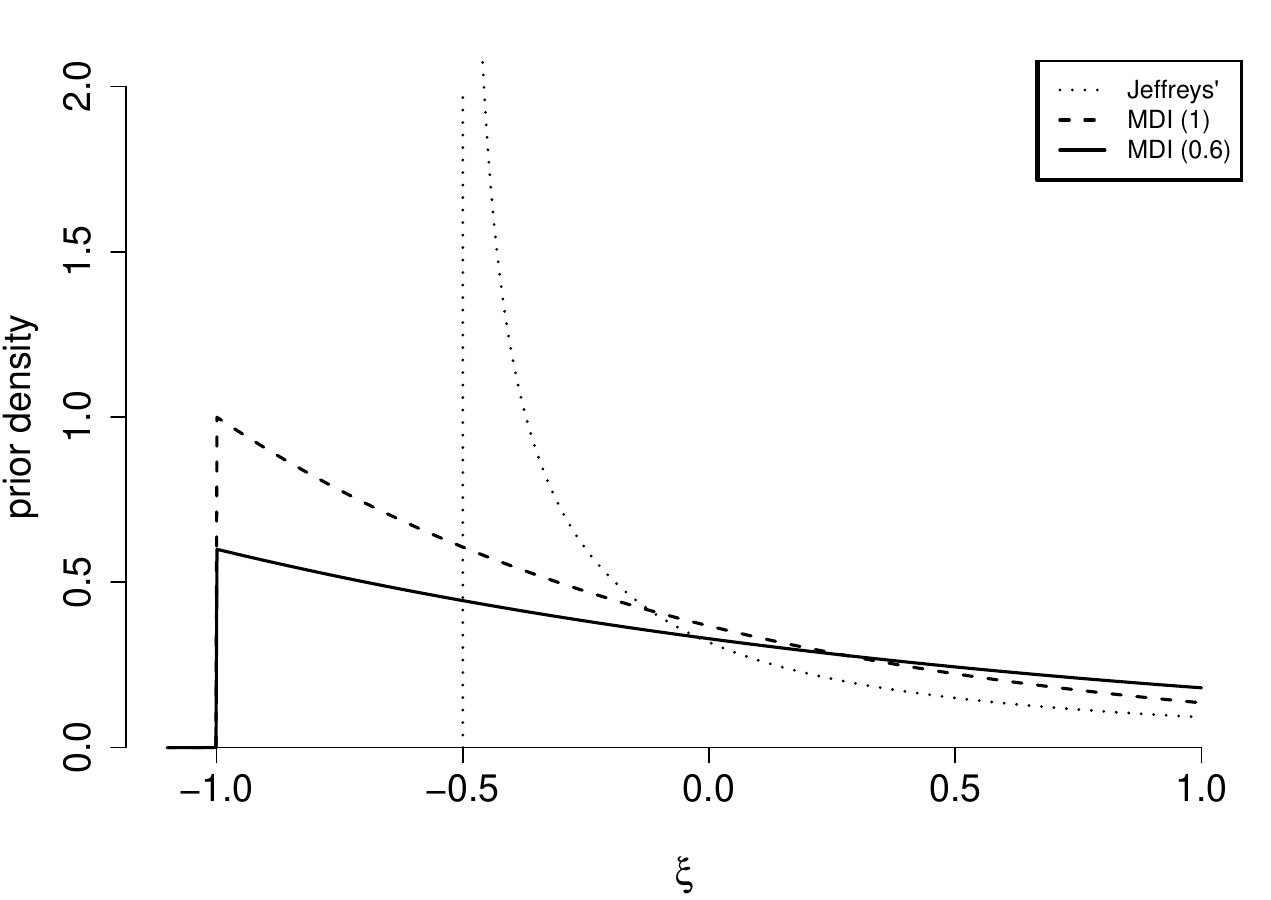}
\vspace{-0.25cm}
\caption{\label{fig:GP_priors} Jeffreys', truncated MDI and generalized MDI priors as functions of $\xi$.}
\end{figure}

We compare approaches for predicting future extreme observations: a predictive approach using different prior distributions and an estimative approach using the MLE.
We use Jeffreys' prior $p_u \sim \mbox{beta}(1/2,1/2)$ for $p_u$, so that
$p_u~|~\bx \sim \mbox{beta}(n_u+1/2, n-n_u+1/2)$, where $n_u$ is the number of threshold excesses. 
Initially we consider three prior distributions for GP parameters: a Jeffreys' prior 
\begin{equation}
\pi_{J}(\sigma_u,\xi) \propto \sigma_u^{-1} (1+\xi)^{-1} (1 + 2 \xi)^{-1/2}, \quad \sigma_u > 0, \,\xi > -1/2; \label{Jeffreys}
\end{equation}
a maximal data information (MDI) prior
\begin{equation}
\pi_{M}(\sigma_u,\xi) \propto \sigma_u^{-1} \,{\rm e}^{-(\xi+1)} \quad \sigma_u > 0, \, \xi \geq -1,  \label{MDI}
\end{equation}
truncated from $\xi \in \mathbb{R}$ to $\xi \geq -1$; and a flat prior 
\begin{equation}
\pi_{F}(\sigma_u,\xi) \propto \sigma_u^{-1}, \qquad \sigma_u>0, \, \xi \in \mathbb{R}. \label{flat}
\end{equation}
Motivated by findings presented later in this section we generalize (\ref{MDI}) to an MDI($a$) prior:
\begin{equation}
\pi_{A}(\sigma,\xi_u;a) \propto \sigma_u^{-1} a\,{\rm e}^{-a(\xi+1)} \quad \sigma_u > 0, \, \xi \geq -1, a > 0. \label{MDIa}
\end{equation}
These priors are improper.  Let $n_u$ be the number of threshold excesses.
\cite{ECC2007} show that the Jeffreys' prior yields a proper posterior for $n_u \geq 1$ and \cite{NA2014} show that under the flat prior a sufficient condition for posterior propriety is $n_u \geq 3$.  
\cite{NA2014} also show that for any sample size, if, and only if, $\xi$ is bounded below {\it a priori}, the MDI prior, and the generalized MDI prior, yield a proper posterior.
At the particular bound of $-1$ used in (\ref{MDI}) the GP distribution reduces to a uniform distribution on $(0,\sigma)$ and corresponds to a change in the behaviour of the GP density: for $\xi < -1$, this density increases without limit as it approaches its mode at the upper end point $-\sigma_u/\xi$, behaviour not expected in extreme value analyses.
Figure \ref{fig:GP_priors} compares the Jeffreys', MDI and generalized MDI prior (for $a=0.6$) as functions of $\xi$.
The Jeffreys' prior (\ref{Jeffreys}) is unbounded as $\xi \downarrow -1/2$.
If there are small numbers of threshold excesses this can result in a bimodal posterior distribution, with one mode at $\xi=-1/2$.
In this simulation study we also find that the Jeffreys' prior results in poorer predictive performance than the truncated MDI and flat priors.

Let $Z_{\rm new}$ be a future $N$-year maximum, sampled from a distribution with distribution function $F(z;\theta)^{n_yN}$.
If the predictive distribution function (\ref{eqn:MNpred}) is the same as that of $Z_{\rm new}$ then $P(M_N \leq Z_{\rm new}~|~\bx)$ has a U(0,1) distribution.  In practice this can only hold approximately: the closeness of the approximation under repeated sampling provides a basis for comparing different prior distributions.  Performance of an estimative approach based on the MLE $\hat{\theta}$ can be assessed using $F(Z_{\rm new};\hat{\theta})^{n_yN}$.
For a given prior distribution and given values of $N, n_y$ and $n$, the simulation scheme is:
\bit
\item [1.] simulate a dataset $\bx_{\rm sim}$ of $n$ independent observations from a BGP($p_u,\sigma_u,\xi$) model and then a sample $\theta_j, j=1, \ldots, m$ from the posterior $\pi(\theta~|~\bx_{\rm sim})$;
\item [2.] simulate an observation $z_{\rm new}$ from the distribution of $M_N$, that is, $\max(X_1, \ldots, X_{N_u})$, where $N_u \sim \mbox{bin}(n_yN,p_u)$ and $X_i \simiid \mbox{GP}(\sigma_u,\xi)$, $i=1, \ldots, N_u$;
\item [3.] use (\ref{eqn:MNpredest}) to evaluate $\hat{P}(M_N \leq z_{\rm new}~|~\bx)$.
\eit
Steps 1. to 3. are repeated 10,000 times, providing a putative sample of size 10,000 from a U$(0,1)$ distribution.  
In the estimative approach step 3 is replaced by evaluation of $F(z_{\rm new};\hat{\theta})^{n_yN}$.
Here, and throughout this paper, we produce samples of size $m$ from the posterior distribution $\pi(\theta~|~\bx)$ using the generalized ratio-of-uniforms method of \cite{WGS1991}, following their suggested strategy of relocating the mode of $\pi(\theta~|~\bx)$ to the origin and setting a tuning parameter $r$ to $1/2$.
In the simulation studies we use $m=1,000$ and when analyzing real data we use $m=10,000$. 

We assess the closeness of the U(0,1) approximation graphically \citep{GA2010},
comparing the proportion of simulated values in each U(0,1) decile to the null value of 0.1.
To aid the assessment of departures from this value we superimpose approximate pointwise 95\% tolerance intervals based on number of points within each decile having a bin($10,000,0.1$) distribution, i.e. $0.1 \pm 1.96\left(0.1 \times 0.9/10,000\right)^{1/2}=0.1 \pm 0.006$.
We use $p_u \in \{0.1, 0.5\}$, $\sigma_u=1$ and values of $\xi$ suggested approximately in section \ref{sec:real1} by the analyses of the Gulf of Mexico data ($\xi \approx 0.1$) and the North Sea data ($\xi \approx -0.2$).

\begin{figure}[H]
\centering
\includegraphics[width=0.9\textwidth, angle=0]{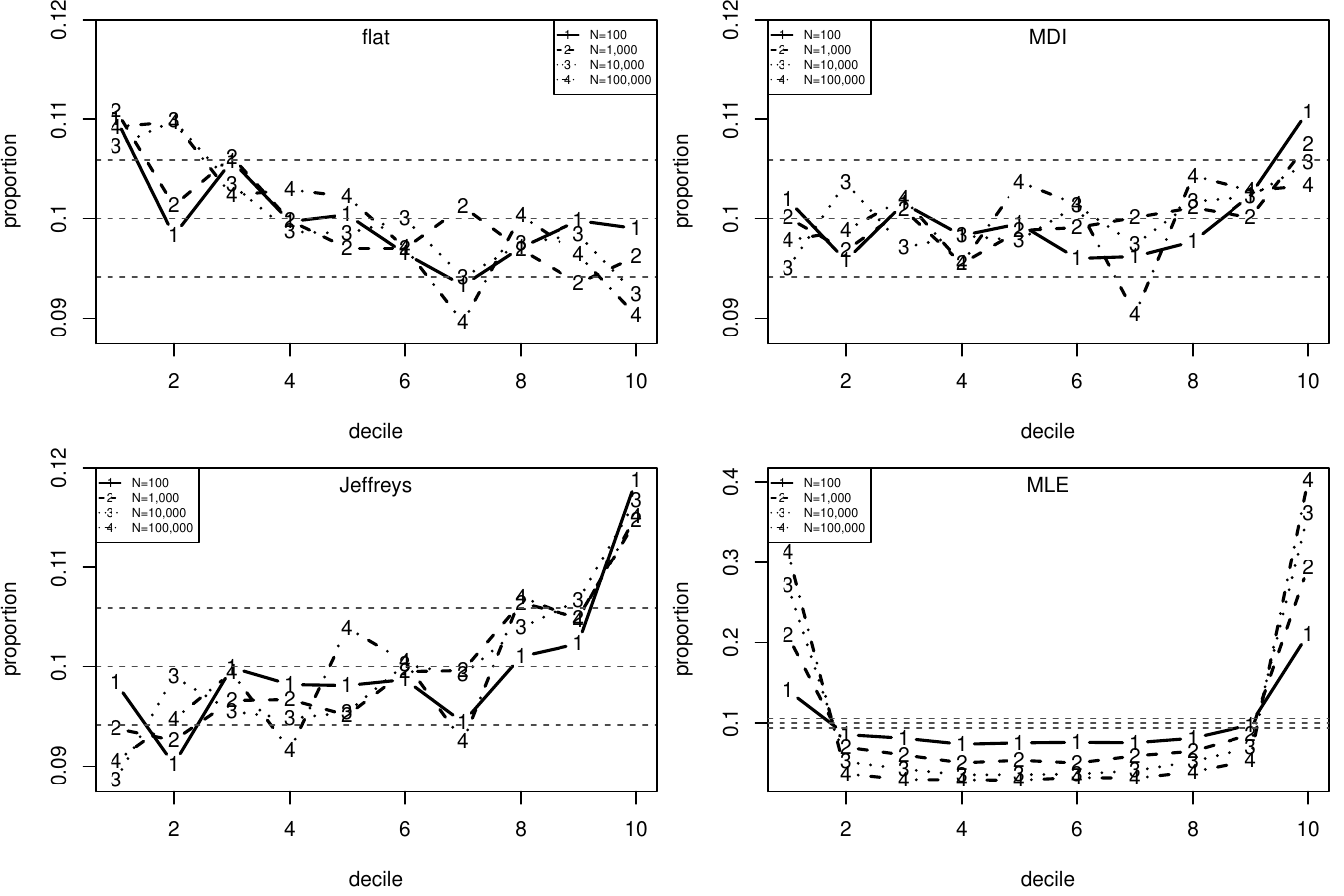}
\vspace{-0.25cm}
\caption{\label{fig:sim1a} Proportions of simulated values of $\hat{P}(M_N \leq z_{\rm new}~|~\bx)$ falling in U(0,1) deciles for the case $\xi=0.1$ and $p_u=0.5$.  The prior is labelled on the plots. Separate lines are drawn for $N=100, 1000, 10000$ and $100000$.   95\% tolerance limits are superimposed.}
\end{figure}

\begin{figure}[H]
\centering
\includegraphics[width=0.9\textwidth, angle=0]{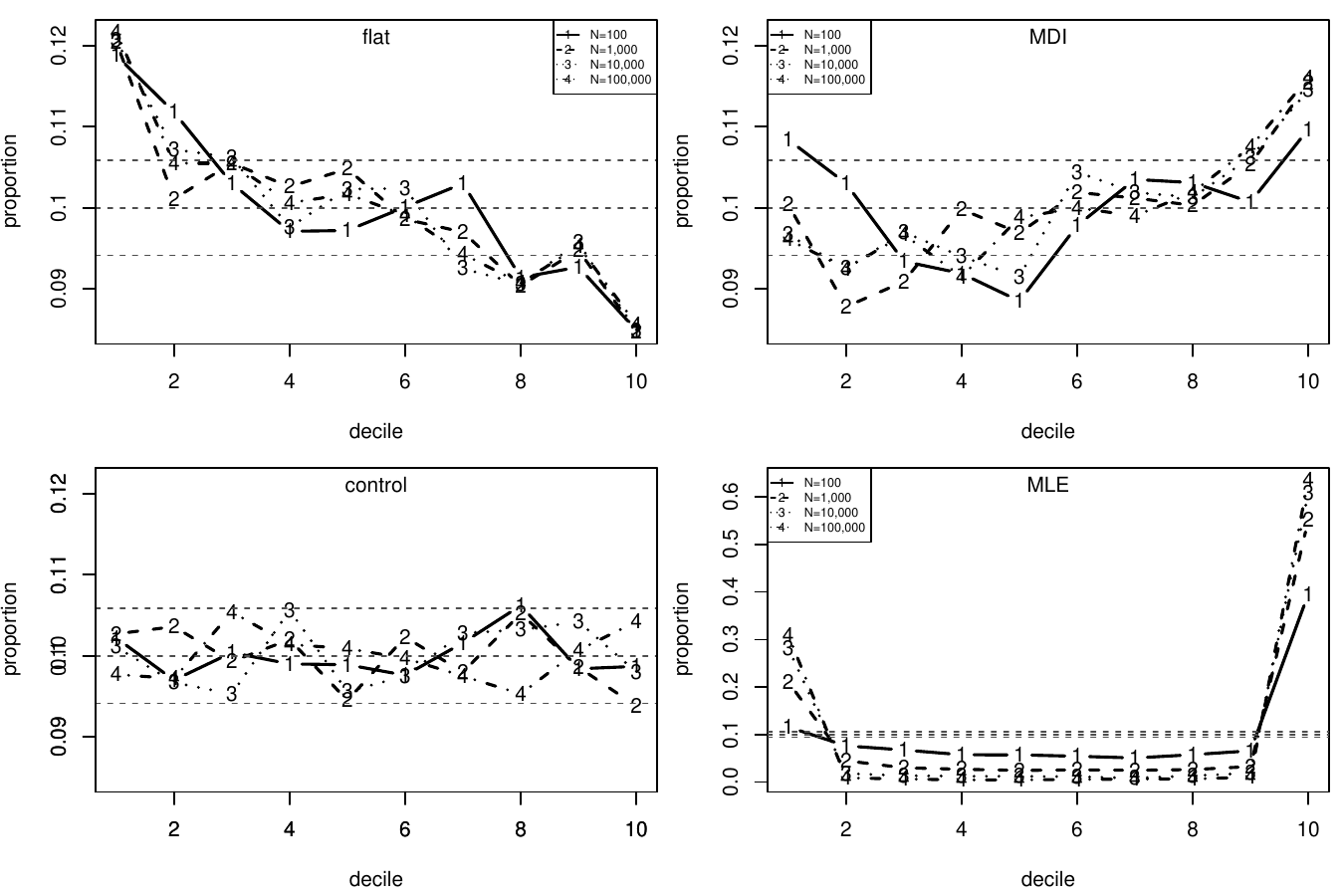}
\vspace{-0.25cm}
\caption{\label{fig:sim1b} Proportions of simulated values of $\hat{P}(M_N \leq z_{\rm new}~|~\bx)$ falling in U(0,1) deciles for the case $\xi=-0.2$ and $p_u=0.1$.  The Jeffreys' prior is replaced by a control plot based on random U(0,1) samples. Separate lines are drawn for $N=100, 1000, 10000$ and $100000$.   95\% tolerance limits are superimposed.}
\end{figure}

The plots in Figures \ref{fig:sim1a} ($\xi=0.1, p_u=0.5$) and \ref{fig:sim1b} ($\xi=-0.2, p_u=0.1$) are based on simulated datasets of length $n=500$ and $n_y=10$, i.e. 50 years of data with a mean of 10 observations per year,  for $N=100, 1000, 10000$ and $100000$ years. 
Note that the plots on the bottom right have much wider $y$-axis scales than the other plots. 
It is evident that the estimative approach based on the MLE produces too few values in deciles 2 to 9 and too many in deciles 1 and 10.  
When the true BGP distribution of $M_N$ (from which $z_{\rm new}$ is simulated in step 2) is wider than that inferred from data (in step 1 and using  (\ref{eqn:MNpredest})) we expect a surplus of values in the first and last deciles.
The estimative approach fails to take account of parameter uncertainty, producing distributions that tend to be too concentrated and resulting in underprediction of large values of $z_{\rm new}$ and overprediction of small values of $z_{\rm new}$.  

The predictive approaches perform much better.
Although departures from desired performance are relatively small, and vary with $N$ in some cases, some general patterns appear.
In Figure \ref{fig:sim1a} the flat prior tends to overpredict large values and small values.  
The MDI prior tends to result in underprediction of large values.  
The Jeffreys prior underpredicts large values, to a greater extent than the MDI prior, and also tends to underpredict small values.
All these tendencies are slightly more pronounced for $\xi=0.1, p_u=0.1$ (not shown).

Figure \ref{fig:sim1b} gives similar findings, although the $N=100$ case behaves a little differently to the larger values of $N$.
The Jeffreys' prior is replaced by a control plot based on values sampled from a U(0,1) distribution.
For $\xi=-0.2$ and with small numbers of threshold excesses the Jeffreys' prior occasionally produces a posterior that is also unbounded as $\xi \downarrow -1/2$, making sampling from the posterior difficult.


\clearpage

These results suggest that, in terms of predicting $M_N$ for large $N$, 
the MDI prior performs better than the flat prior and the Jeffreys' prior.
However, a prior for $\xi$ that is in some sense intermediate between the flat prior and the MDI prior could possess better properties.  To explore this we consider the prior (\ref{MDIa}) for $0 < a \leq 1$.  Letting $a \rightarrow 0$ produces flat prior for $\xi$ on the interval $[-1,\infty)$.  
In order to explore quickly a range of values for $a$ we reuse the posterior samples
based on the priors $\pi_F(\sigma,\xi)$ and $\pi_M(\sigma,\xi)$.
We use the importance sampling ratio estimator (\ref{eqn:imp}) to estimate 
$P(M_N \leq Z_{\rm new}~|~\bx)$ twice, once using $\pi_F(\theta~|~\bx)$ as the importance sampling density $h(\theta)$ and once using $\pi_M(\theta~|~\bx)$.
We calculate an overall estimate of $P(M_N \leq Z_{\rm new}~|~\bx)$ using a weighted mean of the two estimates, with weights equal to the reciprocal of the estimated variances of the estimators \citep[page 603]{Davison2003}.

\begin{figure}[h]
\centering
\includegraphics[width=0.9\textwidth, angle=0]{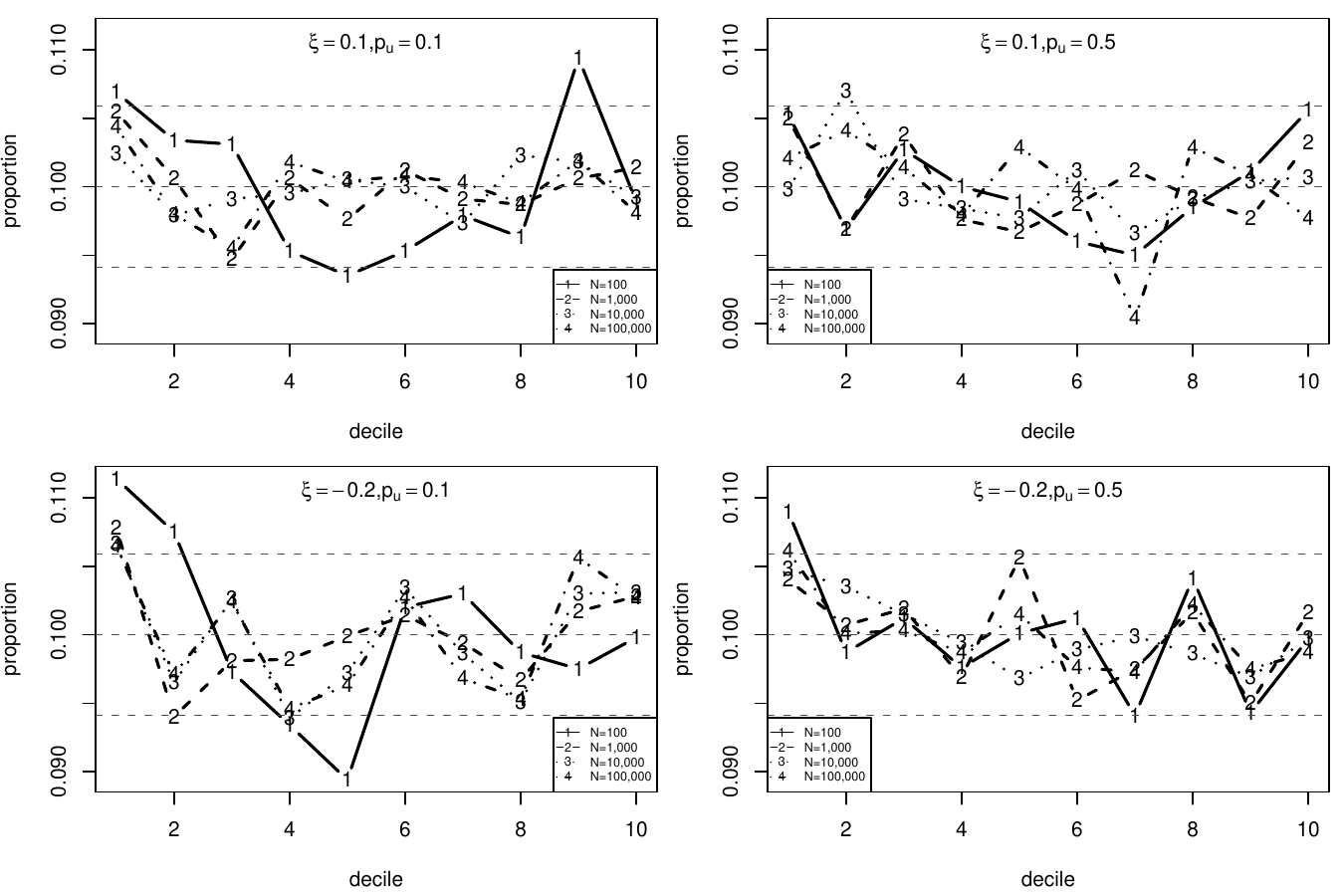}
\vspace{-0.25cm}
\caption{\label{fig:sim1c} Proportions of simulated values of $\hat{P}(M_N \leq z_{\rm new}~|~\bx)$ falling in U(0,1) deciles for different combinations of $\xi$ and $p_u$ under the MDI(0.6) prior.  Separate lines are drawn for $N=100, 1000, 10000$ and $100000$.   95\% tolerance limits are superimposed. }
\end{figure}

Figure \ref{fig:sim1c} shows plots based on the MDI(0.6) prior.  This value of $a$ has been selected based on plots for $a \in \{ 0.1, 0.2, \ldots, 0.9\}$.   We make no claim that this is optimal, just that it is a reasonable compromise between the flat and MDI priors, providing relatively good predictive properties for the cases we have considered.

\subsection{Significant wave height data: single thresholds}
\label{sec:real1}
We analyse the North Sea and Gulf of Mexico storm peak significant wave heights using the MDI(0.6) prior suggested by the simulation study in section \ref{sec:sim1}.  
We use the methodology proposed in section \ref{sec:CVthresh} to quantify the performance of different training thresholds.
We use {\it training thresholds} set at the (0, 5, \ldots, $u_{k}$)\% sample quantiles, for different $u_{k}$.
We define the estimated {\it threshold weight} associated with training threshold $u_i$, assessed at {\it validation threshold} $v(=u_k)$, by
\begin{equation}
w_i(v)=\exp\{\hat{T}_v(u_i)\} / \sum_{j=1}^k \exp\{\hat{T}_v(u_j)\}, \label{eqn:tw2}
\end{equation}
where $\hat{T}_v(u)$ is defined in (\ref{eqn:tw}).
The ratio $w_2(v)/w_1(v)$, an estimate of a {\it pseudo-Bayes factor} \citep{GE1979}, is a measure of the relative performance of threshold $u_2$ compared to threshold $u_1$.
In section \ref{sec:BMA} these weights will be used to combine inferences from different training thresholds.

\begin{figure}[h]
\centering
\includegraphics[width=0.95\textwidth, angle=0]{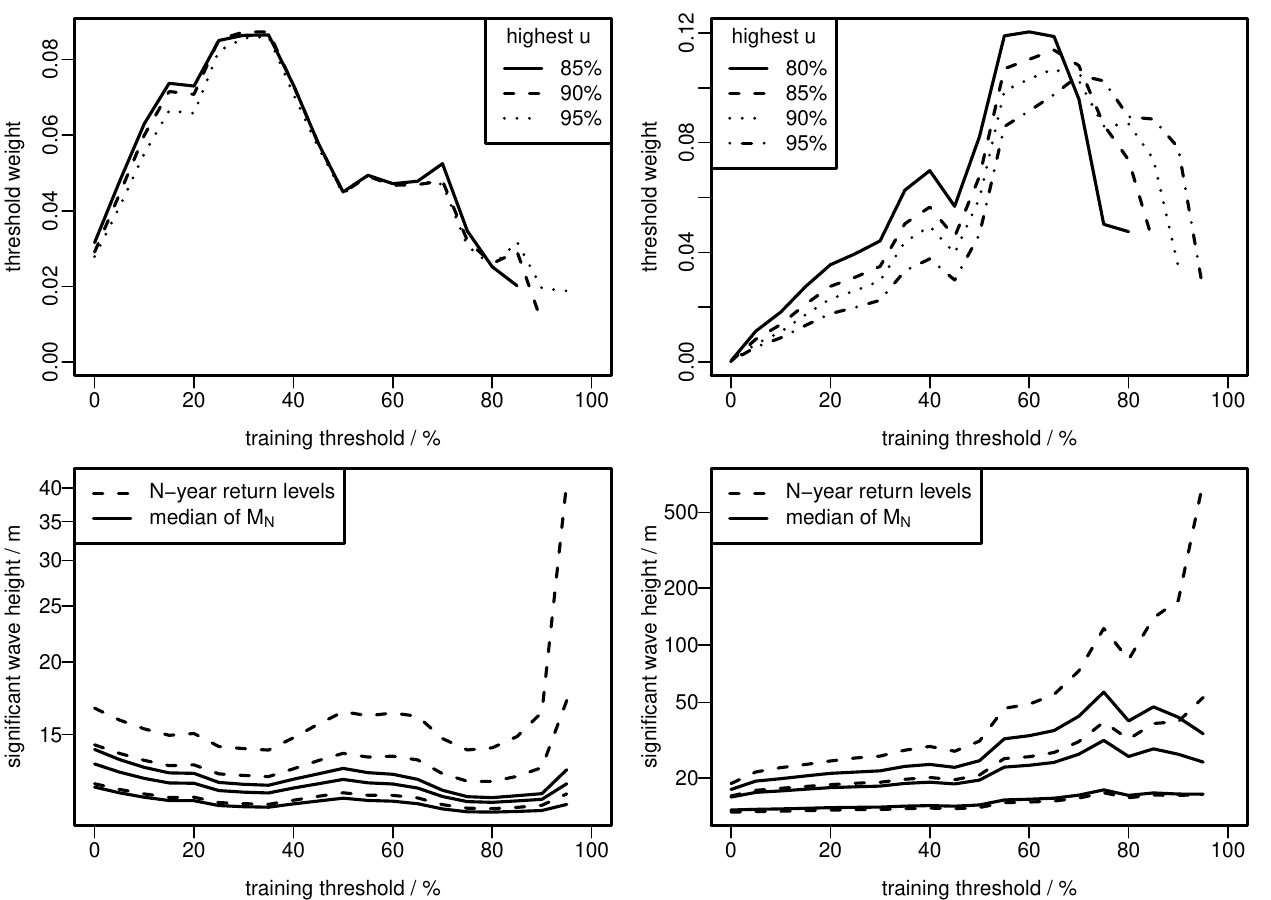}
\vspace{-0.1cm}
\caption{\label{fig:weights}
Analyses of significant wave height data by training threshold $u$.
Top row: estimated threshold weights by the highest training threshold considered.
Bottom row: $N$-year predictive return levels and medians of the predictive distribution of $M_N$ for $N=100, 1000$ and $10000$.
Left: North Sea.  Right: Gulf of Mexico.}
\end{figure}

The top row of Figure \ref{fig:weights} shows plots of the estimated training weights against training threshold based for different $u_{k}$.
For the North Sea data training thresholds in the region of the sample 25-35\% quantiles 
(for which the MLE of $\xi \approx -0.2$) have relatively large threshold weight and there is little sensitivity to $u_{k}$.
For the Gulf of Mexico data training thresholds in the region of the 60-70\% sample quantiles (for which the MLE of $\xi \approx 0.1$) are suggested, and the threshold at which the largest weight is attained is more sensitive to $u_{k}$.
As expected from the histogram of the Gulf of Mexico data in Figure \ref{fig:Hs_data} training thresholds below the 25\% sample quantile have low threshold weight. 
Note that for the Gulf of Mexico data the 90\% and 95\% thresholds have far fewer excesses (32 and 16) than the suggested 50.

\begin{figure}[h]
\centering
\includegraphics[width=0.75\textwidth, angle=0]{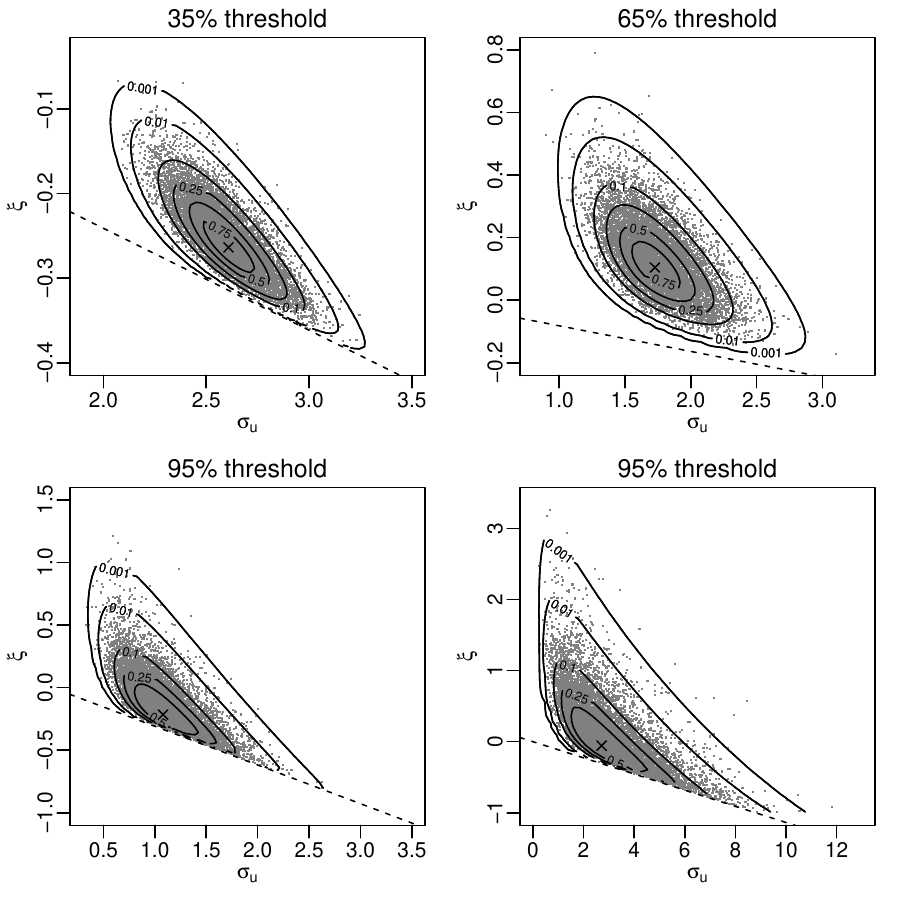}
\vspace{-0.25cm}
\caption{\label{fig:posteriors} Samples from $\pi(\sigma_u, \xi~|~\bx)$, with density contours scaled to unity at the posterior mode ($\times$).   Dashed lines show the support of the posterior distribution: $\xi>\sigma_u/(x_n-u)$.
Top: `best' training threshold (for $u_{k}$ set at the 85\% sample quantile).  Bottom: 95\% training threshold.  
Left: North Sea.  Right: Gulf of Mexico.}
\end{figure}

The bottom row of Figure \ref{fig:weights} shows that the $N$-year predictive return levels and the medians of the predictive distribution of $N$-year maxima $M_N$ are close for $N=100$, where little or no extrapolation is required, but for $N=1000$ and $N=10000$ the former is much greater than the latter.  
For the North Sea data the results appear sensible and broadly consistent with estimates from elsewhere.
From the 55\% training threshold upwards, which includes thresholds that have high estimated training weights, estimates of the median of $M_{1000}$ and $M_{10000}$ from the Gulf of Mexico data are implausibly large, e.g. 31.6m and 56.7m for the 75\% threshold.
The corresponding estimates of the predictive return levels are even less credible.
The problem is that high posterior probability of large positive values of $\xi$, caused by high posterior uncertainty about $\xi$, translates into large predictive estimates of extreme quantiles.  
Figure \ref{fig:posteriors} gives examples of the posterior samples of $\sigma_u$ and $\xi$ underlying the plots in Figure \ref{fig:weights}.
The marginal posterior distributions of $\xi$ are positively skewed, particularly so for the 95\% training thresholds, mainly because for fixed $\sigma_u$, $\xi$ is bounded below by $\sigma_u/(x_n-u)$.
The higher the threshold the larger the posterior uncertainty and the greater the skewness towards values of $\xi$ that correspond to a heavy-tailed distribution.  For the Gulf of Mexico data at the 95\% threshold $\hat{P}(\xi>1/2 \mid \bm{x}) \approx 0.20$ and $\hat{P}(\xi>1 \mid \bm{x}) \approx 0.05$.
This issue is not peculiar to a Bayesian analysis: frequentist confidence intervals for $\xi$ and for extreme quantiles are also unrealistically wide.  


Physical considerations suggest that there is a finite upper limit to storm peak $H_s$ \citep{JE2013}.
However, if there is positive posterior probability on $\xi\geq 0$ then the implied distribution of $H_s$ is unbounded above and
on extrapolation to a sufficiently long time horizon, $N_l$ say, unealistically large values will be implied.
This may not be a problem if $N_l$ is greater than the time horizon of practical interest, that is,
the information in the data is sufficient to allow extrapolation over this time horizon.
If this is not the case then one should incorporate supplementary data (perhaps by pooling data over space as in \cite{NJ2011}) or prior information.  
Some practitioners assume that $\xi < 0$ {\it a priori}, in order to ensure a finite upper limit, but
such a strategy may sacrifice performance at time horizons of importance
and produce unrealistically small estimates for the magnitudes of rare events.
In section \ref{sec:WIprior} we consider how one might specify a prior for $\xi$ that allows the data to dominate 
when it contains sufficient information to do so and avoids unrealistic inferences otherwise.

\section{Accounting for uncertainty in threshold}
\label{sec:BMA}
We use Bayesian model-averaging \citep{HMRV1999,GD1994} to combine inferences based on different thresholds.
Consider a set of $k$ training thresholds $u_1, \ldots, u_k$ and a particular validation threshold $v$.
We view the $k$ BGP models associated with these thresholds as competing models.
There is evidence that one tends to get better predictive performance by interpolating smoothly between all models entertained as plausible {\it a priori}, than by choosing a single model \citep[section 7]{HMRV1999}.
Suppose that we specify prior probabilities $P(u_i), i=1, \ldots, k$ for these models.
In the absence of more specific prior information, and in common with \cite{WT2012},  we use a discrete uniform prior $P(u_i)=1/k, i=1, \ldots,k$.
We suppose that the thresholds occur at quantiles that are equally spaced on the probability scale.  We prefer this to equal spacing on the data scale because it seems more natural than an equal spacing on the data scale and retains its property of equal spacing under data transformation.

Let $\theta_i=(p_i,\sigma_i,\xi_i)$ be the BGP parameter vector under model $u_i$, under which the prior is $\pi(\theta_i~|~u_i)$.
By Bayes' theorem, the {\it posterior threshold weights} are given by
\[ P_v(u_i~|~\bx) = \frac{f_v(\bx~|~u_i)\,P(u_i)}{\sum_{j=1}^k f_v(\bx~|~u_j)\,P(u_j)}, \]
where 
$f_v(\bx~|~u_i) = \int f_v(\bx~|~\theta_i, u_i) \pi(\theta_i~|~u_i) {\rm ~d}\theta_i$
is the prior predictive density of $\bx$ based on validation threshold $v$ under model $u_i$.
However, $f_v(\bx~|~u_i)$ is difficult to estimate and is improper if $\pi(\theta_i~|~u_i)$ is improper.
Following \cite{GE1979} we use 
$\prod_{r=1}^n f_v(x_r~|~\bx_{(r)}, u_i)=\exp\{\hat{T}_v(u_i)\}$ 
as a surrogate for $f_v(\bx~|~u_i)$ to give
\begin{equation}
\hat{P}_v(u_i~|~\bx) = \frac{\exp\{\hat{T}_v(u_i)\}\,P(u_i)}
{\sum_{j=1}^k \exp\{\hat{T}_v(u_j)\}\,P(u_j)}.  \label{eqn:weights}
\end{equation}

Let $\theta_{ij}, j=1, \ldots, m$ be a sample from $\pi(\theta_i~|~\bx)$,
the posterior distribution of the GP parameters based on threshold $u_i$.
We calculate a threshold-averaged estimate of the predictive distribution function of $M_N$ using
\begin{equation}
\hat{P}_v(M_N \leq z~|~\bx)= \sum_{i=1}^k \hat{P}(M_N \leq z~|~\bx, u_i) 
\hat{P}_v(u_i~|~\bx), \label{eqn:mod_ave}
\end{equation}
where, by analogy with (\ref{eqn:MNpredest}),
$\hat{P}(M_N \leq z~|~\bx, u_i) = (1/m)\sum_{j=1}^m F(z;\theta_{ij})^{n_yN}$.
The solution $\hat{z}_{PM}(N)$ of
\begin{equation}
\hat{P}_v(M_1 \leq \hat{z}_{PM}(N)~|~\bx) =1-1/N \label{eqn:ret_lev_BMA}
\end{equation}
provides a threshold-averaged estimate of the $N$-year predictive return level, based on validation threshold $v$.

\subsection{Simulation study 2: single and multiple thresholds}
\label{sec:sim2}
We compare inferences from a single threshold to those from averaging over many thresholds,
based on random samples simulated from three distributions, chosen to represent qualitatively different behaviours.  
With knowledge of the simulation model we should be able to choose a suitable single threshold, at least approximately.  
In practice this would not be the case and so it is interesting to see how well the strategies of choosing the `best' threshold $u^*$ (section \ref{sec:CV}), and of averaging inferences over different thresholds (section \ref{sec:BMA}), compare to this choice and how the estimated weights $\hat{P}_v(u_i~|~\bx)$ in (\ref{eqn:weights}) vary over $u_i$.

The three distributions are now described.  
A (unit) exponential distribution has the property that a GP(1,0) model holds above any threshold.  
Therefore, choosing the lowest available threshold is optimal.  
For a (standard) normal distribution the GP model does not hold for any finite threshold, the quality of a GP approximation improving slowly as the threshold increases.  
In the limit $\xi=0$, but at finite levels the effective shape parameter is negative \citep{WT2012} and one expects a relatively high threshold to be indicated.  
A uniform-GP hybrid has a constant density up to its 75\% quantile and a GP density (here with $\xi=0.1$) for excesses of the 75\% quantile.  
Thus, a GP distribution holds only above the 75\% threshold.

In each case we simulate 1000 samples each of size 500, representing 50 years of data with an average of 10 observations per year.  
We set training thresholds at the $50\%, 55\%, \ldots, 90\%$ sample quantiles, so that there are 50 excesses of the
(90\%) validation threshold.
For each sample, and for values of $N$ between $100$ and $10,000$, we solve 
$\hat{P}_v(M_N \leq z~|~\bx)=1/2$ for $z$ (see  (\ref{eqn:MNpredest})) to give estimates of the median of $M_N$.  We show results for three single thresholds: the threshold one might choose based on knowledge of the simulation model; the `best' threshold $u^*$ (see section \ref{sec:CV}); and another (clearly sub-optimal) threshold chosen to facilitate further comparisons.  
We compare these estimates, and a threshold-averaged estimate based on (\ref{eqn:mod_ave}) to the true median of $M_N$, $H^{-1}((1/2)^{10N})$,
where $H$ is the distribution function of the underlying simulation model.

\begin{figure}[p]
\centering
\includegraphics[width=0.85\textwidth, angle=0]{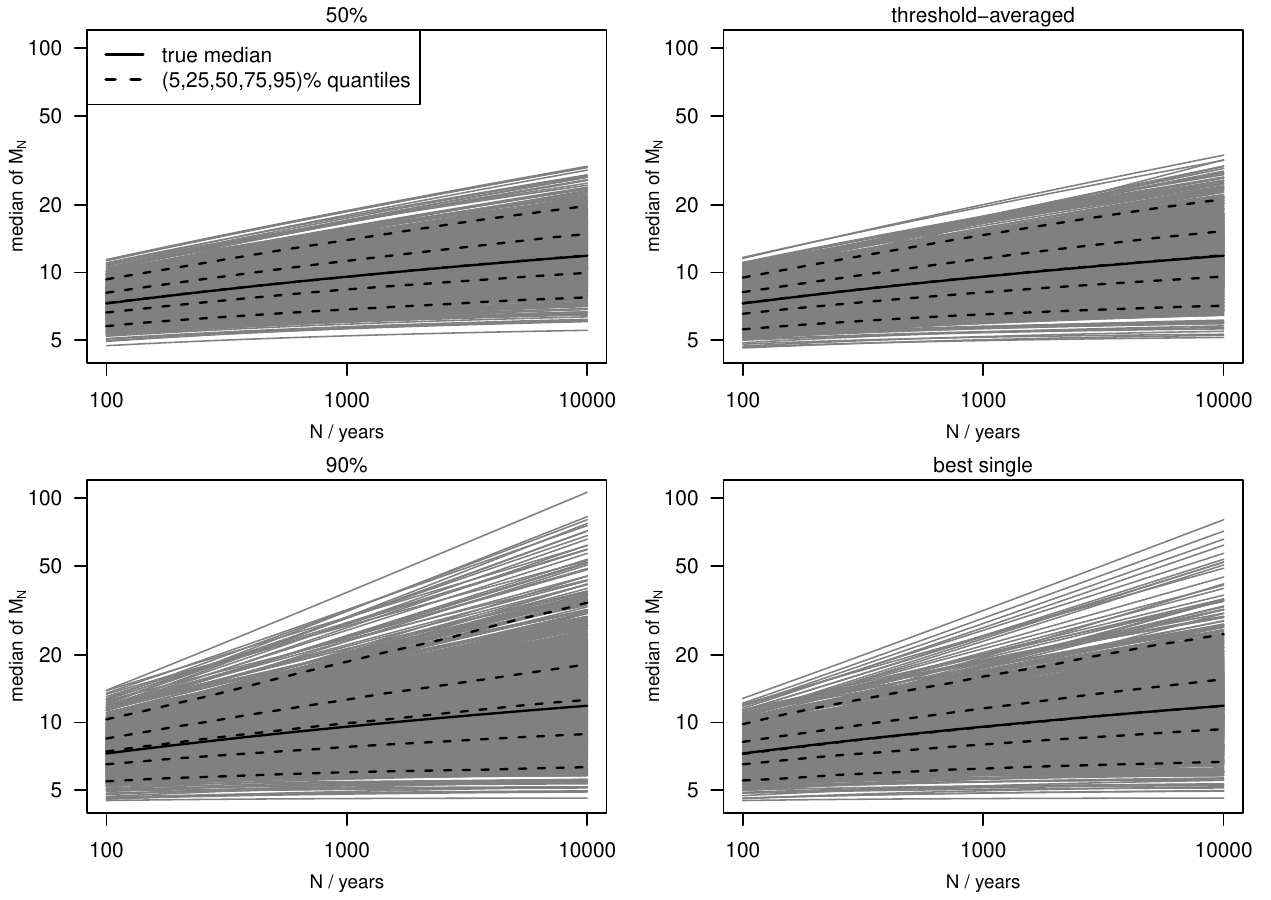}
\vspace{-0.25cm}
\caption{\label{fig:sim2_exp} 
Exponential example.  
Predictive median of $M_N$ by $N$: individual datasets (grey), $N$-specific (5,25,50,75,95)\% sample quantiles (dashed) and true median (solid).  
Threshold strategies: median (top left); 90\% quantile (bottom left);
threshold-averaged (top right); `best' threshold (bottom right).
}
\end{figure}

\begin{figure}[p]
\centering
\includegraphics[width=0.85\textwidth, angle=0]{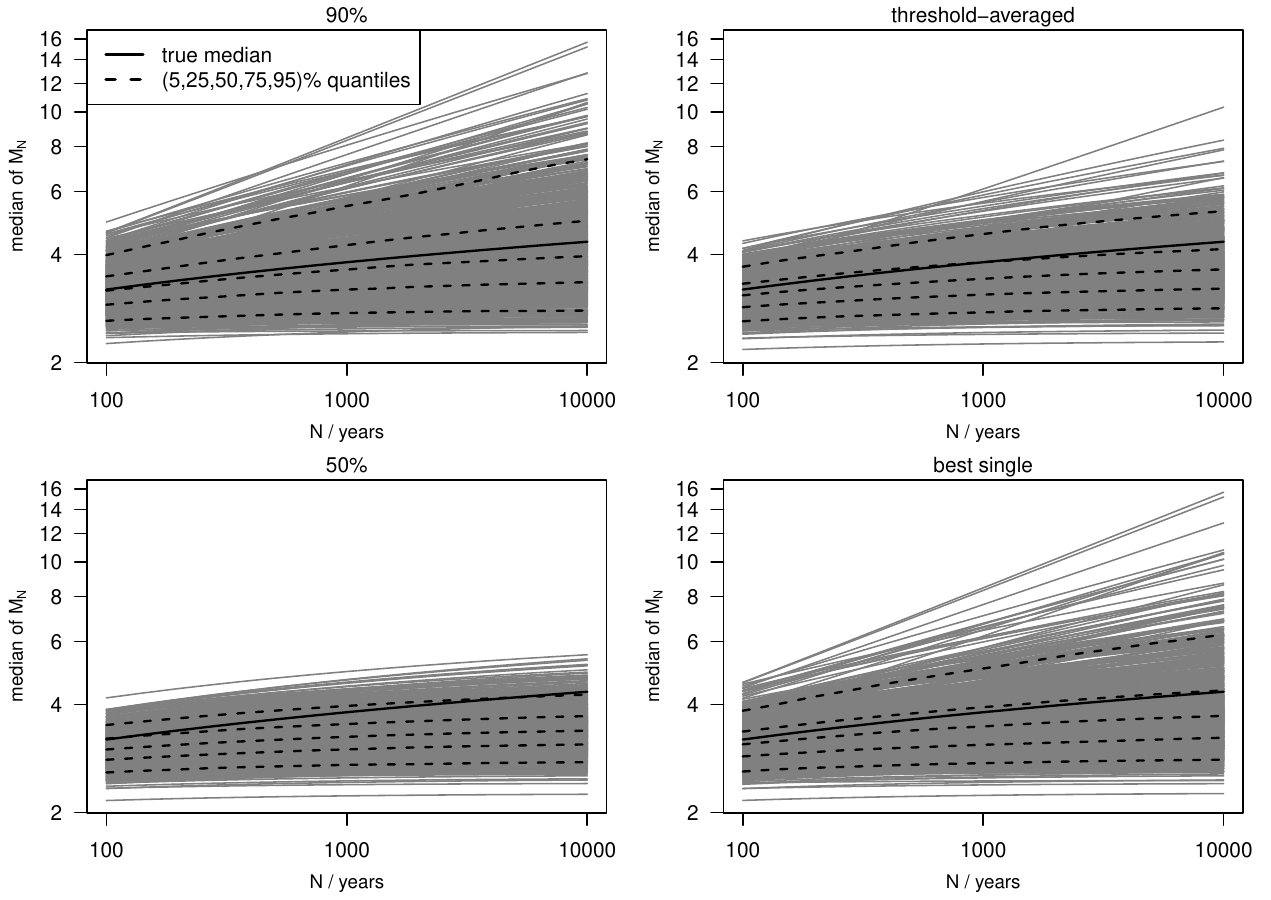}
\vspace{-0.25cm}
\caption{\label{fig:sim2_norm} 
Normal example.  
Predictive median of $M_N$ by $N$: individual datasets (grey), $N$-specific (5,25,50,75,95)\% sample quantiles (dashed) and true median (solid).  
Threshold strategies: 90\% quantile (top left); median (bottom left);
threshold-averaged (top right); `best' threshold (bottom right).
}
\end{figure}

\begin{figure}[p]
\centering
\includegraphics[width=0.85\textwidth, angle=0]{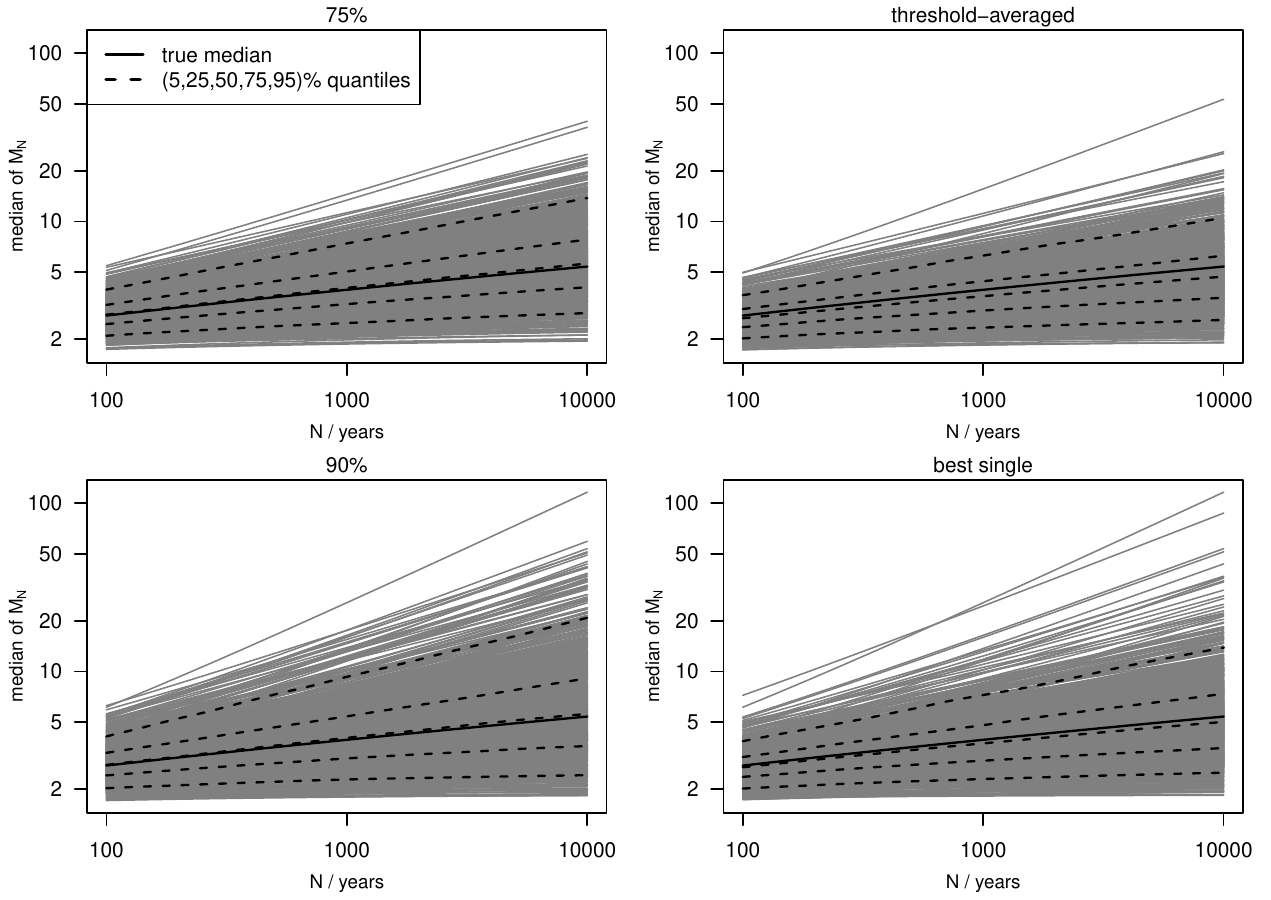}
\vspace{-0.25cm}
\caption{\label{fig:sim2_hybrid} 
Uniform-GP hybrid example.  
Predictive median of $M_N$ by $N$: individual datasets (grey), $N$-specific (5,25,50,75,95)\% sample quantiles (dashed) and true median (solid).  
Threshold strategies: 75\% quantile (top left); 90\% quantile (bottom left);
threshold-averaged (top right); `best' threshold (bottom right).
}
\end{figure}

The results for the exponential distribution are summarized in Figure \ref{fig:sim2_exp}.  As expected, all strategies have negligible bias.   
The threshold-averaged estimates match closely the behaviour of the optimal strategy (the 50\% threshold).  
The best single threshold results in slightly greater variability, offering less protection than threshold-averaging against estimates that are far from the truth.
In the normal case (Figure \ref{fig:sim2_norm}) the expected underestimation is evident for large $N$: this is substantial for a 50\% threshold but small for a 90\% threshold.  
The CV-based strategies have greater bias than those based on a 90\% threshold, because inferences from lower thresholds contribute, but have much smaller variability.  
Similar findings are evident in Figure \ref{fig:sim2_hybrid} for the uniform-GP hybrid distribution: contributions from thresholds lower than the 75\% quantile produce negative bias but threshold-averaging achieves lower variability than the optimal 75\% threshold.

In all these examples the CV-based strategies seem preferable to a poor choice of a single threshold, and, in a simple visual comparison of bias and variability, are not dominated clearly by a (practically unobtainable) optimal threshold.  
Using threshold-averaging to account for threshold uncertainty is conceptually attractive but, the exponential example aside, compared to the `best' threshold strategy it's reduction in variability is at the expense of slightly greater bias.   
A more definitive comparison would depend on problem-dependent losses associated with over- and under-estimation.  

\begin{figure}[p]
\centering
\includegraphics[width=0.85\textwidth, angle=0]{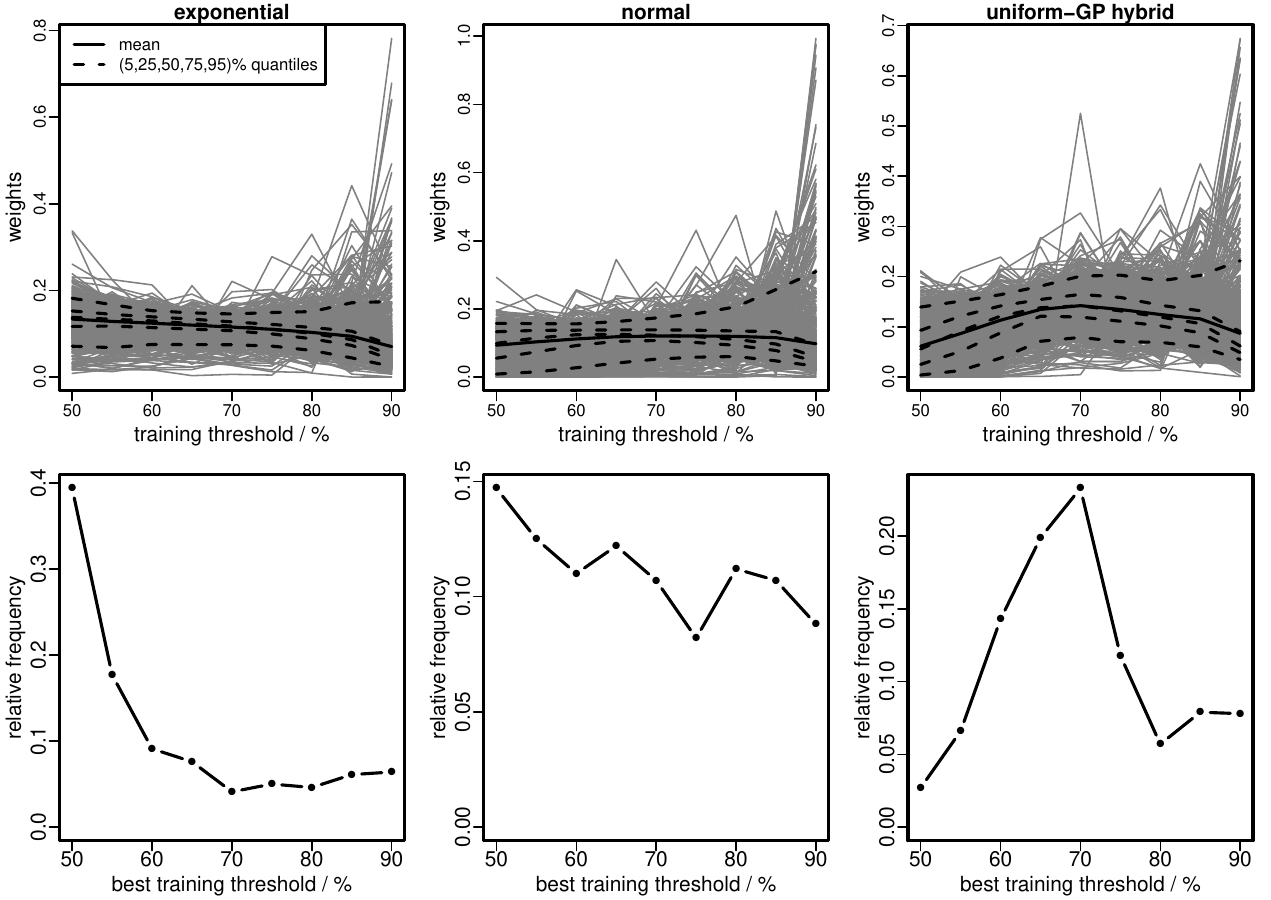}
\vspace{-0.25cm}
\caption{\label{fig:sim2_weights} Threshold weights by training threshold.
Top: individual datasets (grey) with threshold-specific sample means (solid black) and (5,25,50,75,95)\% sample quantiles (dashed).
Bottom: relative frequency with which each threshold has the largest weight. 
Left: exponential.  Middle: normal. Right: uniform-GP hybrid.}
\end{figure}

Figure \ref{fig:sim2_weights} summarizes how the posterior threshold weights vary with training threshold. 
For a few datasets the 90\% training threshold receives highest weight.  
This occurs when inferences about $\xi$ using a 90\% threshold differ from those using each lower threshold.
This effect is diminishes if the number of excesses in the validation set is increased.  
In the exponential and hybrid cases the average weights behave as expected: decreasing in $u$ in the exponential case, and peaking at approximately the 70\% quantile (i.e. lower than the 75\% quantile) in the uniform-GP case. 
In the exponential example the best available threshold (the 50\% quantile) receives the highest weight with relatively high probability.
In the hybrid example the 70\% quantile receives the highest weight most often.
The 75\% quantile is the lowest threshold at which the GP model for threshold excesses is correct.
The 70\% quantile performs better than the 75\% quantile by trading some model mis-specification bias for increased precision resulting from larger numbers of threshold excesses.
In the normal case there is no clear-cut optimal threshold.
This is reflected in the relative flatness of the graphs, with the average weights peaking at approximately the 70-80\% quantile and the 50\% threshold being the `best' slightly more often than higher thresholds. 
Given the slow convergence in this case it may be that much higher thresholds should be explored, requiring much larger simulated sample sizes, such as those used by \cite{WT2012}.

\clearpage

\subsection{Significant wave height data: threshold uncertainty}
\label{sec:real2}
\begin{figure}[h]
\centering
\includegraphics[width=0.9\textwidth, angle=0]{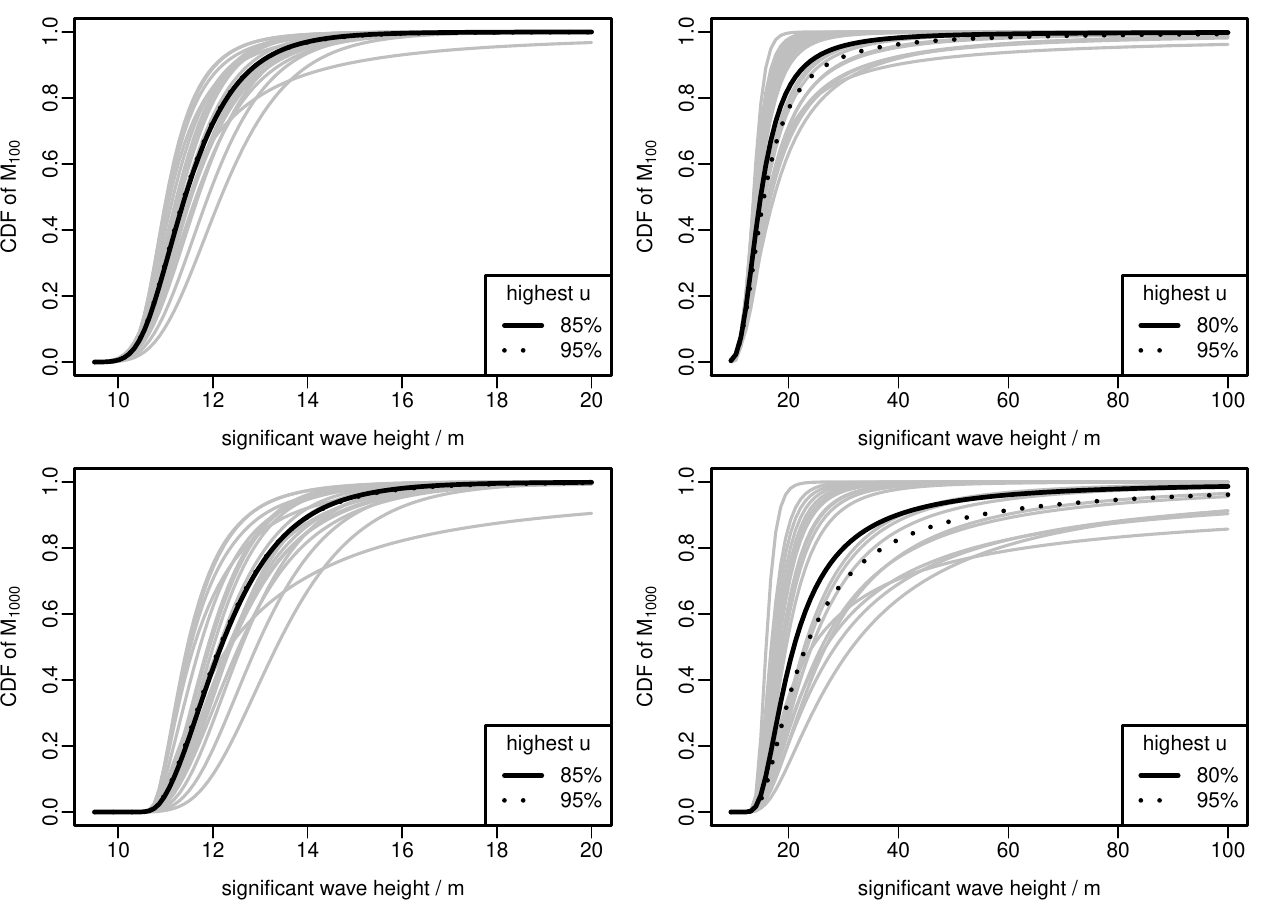}
\vspace{-0.25cm}
\caption{\label{fig:BMA} Threshold-specific (grey lines) and threshold-averaged (black lines) predictive distribution functions of $M_{100}$ (top) and $M_{1000}$ (bottom).   
Left: North Sea data. Right: Gulf of Mexico data.}
\end{figure}

We return to the significant wave height datasets, using the methodology of section \ref{sec:BMA} to average extreme value inferences obtained from different thresholds.
Figure \ref{fig:BMA} shows the estimated threshold-specific predictive distribution functions of $M_{100}$ and $M_{1000}$.
Also plotted are estimates from the weighted average (\ref{eqn:mod_ave}) over thresholds, for different choices of the highest threshold $u_k$.
For the North Sea data there is so little sensitivity to $u_k$ that the black curves are indistinguishable.
For the Gulf of Mexico data there is greater sensitivity to $u_k$, although based on the discussion in section \ref{comp_thresh} setting $u_k$ at the 95\% sample quantile is probably inadvisable with only 315 observations. 
However, for both choices of $u_k$, averaging inferences over thresholds has provided some protection against the high probability of unrealistically large values of $H_s$ estimated under some individual thresholds.  

\subsection{A weakly-informative prior for \bm{$\xi$}}
\label{sec:WIprior}
We have used prior distributions for model parameters that are constructed without reference to the particular problem in hand.
This strategy is inadvisable when the data contain insufficient information to dominate such priors, because inferences are then influenced strongly by a generically-chosen prior.
When analysing the Gulf of Mexico data in section \ref{sec:real1} we found that for the highest thresholds there was insufficient information in the data to avoid unrealistic extreme value extrapolations at long time horizons.
If small sample sizes cannot be avoided and long time horizons are important then unrealistic inferences can be avoided by providing application-specific prior information.
This prior could be elicited from an expert, as in \cite{CT1996}, or specified to reflect general experience of the quantity under study, such as the beta-type prior for $\xi$ on $-1/2 \leq \xi \leq 1/2$ used by \cite{Martins2001} for river flows and rainfall totals.

Here, we illustrate the effects on the Gulf of Mexico analysis of providing weak information about $\xi$, with the aim of preventing unrealistic inferences rather than specifying strong prior information. 
The particular prior we use is 
\begin{equation}
\pi_{C}(\sigma_u,\xi;A) \propto \sigma_u^{-1} \, (1+\xi^2/A^2)^{-1}  \quad \sigma_u > 0, \, \xi \geq -1, A>0, \label{Cauchy}
\end{equation}
so that the marginal prior for $\xi$ is a Cauchy distribution, truncated to $\xi \geq -1$.
The prior mode is zero, reflecting the expectation that $\xi$ somewhat close to zero.
The use of a Cauchy density is motivated by \cite{Gelman2006}, who uses a half-Cauchy prior for a group-level standard deviation in a hierarchical model.  
The scale parameter $A$ is set, guided by expert information from oceanographers (see below), to downweight {\it a priori} unrealistically large values of $\xi$.
The tails of a Cauchy density have a gentle slope, so information from the data can be influential if it conflicts with this prior.  

Let $m_N$ be the median of $M_N$.  
Oceanographers with expert knowledge of the hurricane-induced storms in the Gulf of Mexico have provided  
approximate values of $\hat{m}_{100}=15$m and 1.5 for the ratio $m_{10000}/m_{100}$.
Estimating $m_1$ directly from the data gives $\hat{m}_1=4.55$m.
With a view to downweighting substantially {\it a apriori} only very unrealistically large values of $\xi$
we set $\hat{m}_{10000}=3\hat{m}_{100}=45$m, that is, we double the ratio provided by the experts.
Based on theory concerning the limiting behaviour of the maximum of i.i.d. random variables \citep[chapter 3]{Coles2001} we suppose that, for $N \geq 1$, $M_N$ has a generalized extreme value GEV($\mu_N, \sigma_N, \xi$) distribution.  
Then $\mu_N=\mu_1+\sigma_1[(N / \ln 2)^{\xi}-1]/\xi$ and the quantity $R=(m_{10000}-m_1)/(m_{100}-m_1)$ is a function only of $\xi$.  
If $\xi=0.229$ then $R=(\hat{m}_{10000}-\hat{m}_1)/(\hat{m}_{100}-\hat{m}_1)$, suggesting a prior for which $P(\xi > 0.229)$ is small.
We use $A=0.154$, for which $P(\xi > 0.229) = 0.2$ and $P(\xi > 1/2)=0.1$.
Note that under the MDI(0.6) prior $P(\xi > 1/2)=0.41$.

\begin{figure}[h]
\centering
\includegraphics[width=0.7\textwidth, angle=0]{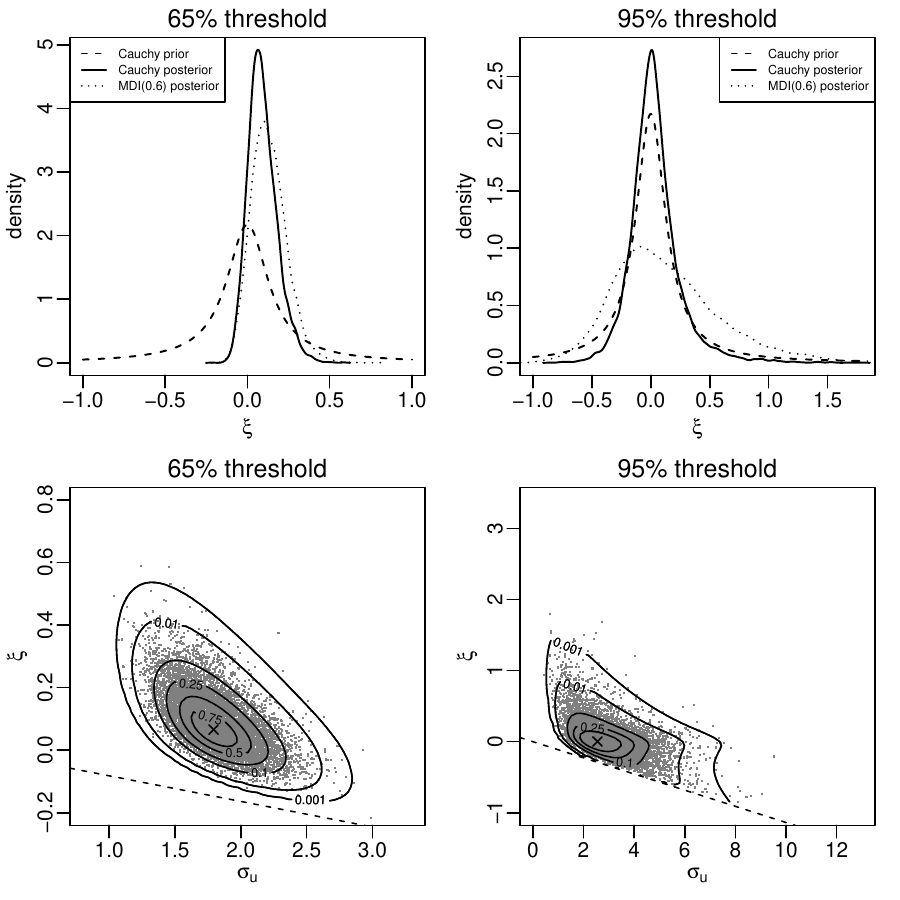}
\vspace{-0.5cm}
\caption{\label{fig:Cposteriors} Posterior samples for the Gulf of Mexico data.  
Top: kernel estimates of the posterior density for $\xi$ under the Cauchy prior (solid) and MDI(0.6) prior (dotted), with the Cauchy prior density (dashed).
Bottom: samples from $\pi(\sigma_u, \xi~|~\bx)$ under the Cauchy prior, with density contours scaled to unity at the mode ($\times$).   Dashed lines show the support of the posterior distribution.
Left: 65\% threshold.  Right: 95\% threshold.}
\end{figure}

Figure \ref{fig:Cposteriors} contains graphical summaries of posterior samples under the Cauchy and MDI(0.6) priors for 65\% and 95\% training thresholds.
For the 65\% threshold, comparing the marginal posterior densities for $\xi$ (top left) shows that the
change of prior has little effect, because at this threshold there is sufficient information in the data to dominate these priors.
Similarly, the plot of the joint posterior (bottom left) is very similar to the corresponding plot in the top right of Figure \ref{fig:posteriors}.
For the 95\% threshold the posterior for $\xi$ under the Cauchy prior is much less diffuse than under the MDI(0.6) prior and far less posterior density is associated with large values of $\xi$.
Compare also the bottom right plots in Figures \ref{fig:posteriors} and \ref{fig:Cposteriors}.
Lack of information the data means that the posterior based on the Cauchy prior is only slightly less diffuse than the prior itself.

\begin{figure}[h]
\centering
\includegraphics[width=0.95\textwidth, angle=0]{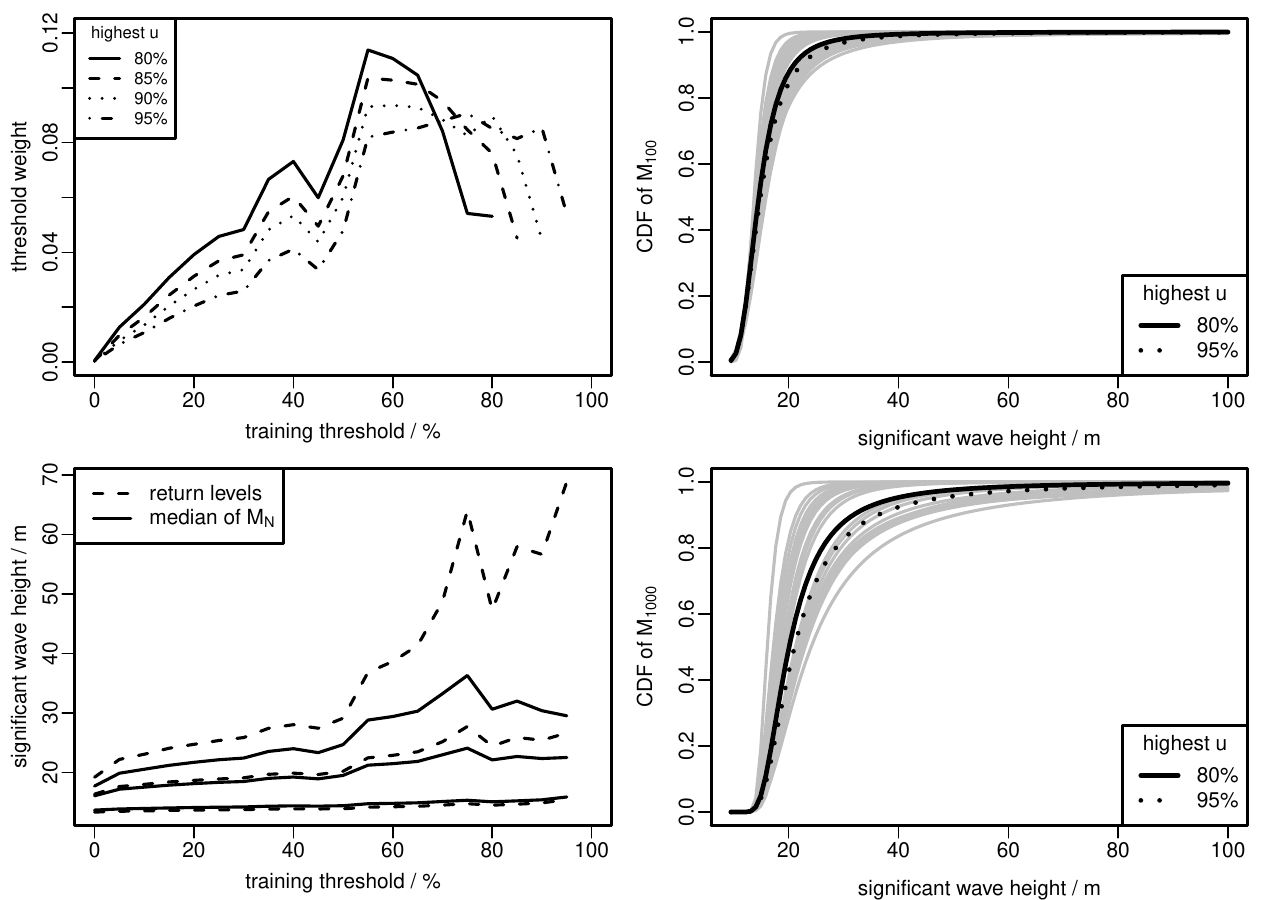}
\vspace{-0.25cm}
\caption{\label{fig:Cthresh} 
Extreme value inferences for the Gulf of Mexico data using a Cauchy prior for $\xi$.
Top left: estimated threshold weights by the highest training threshold.
Bottom left: $N$-year predictive return levels and medians of the predictive distribution of $M_N$ for $N=100, 1000$ and $10000$.
Right: threshold-specific (grey lines) and threshold-averaged (black lines) predictive distribution functions of $M_{100}$ (top) and $M_{1000}$ (bottom).}
\end{figure}

The plots on the left of Figure \ref{fig:Cthresh} show, by comparison with the plots on the right of Figure \ref{fig:weights}, the effect of the change of prior on the threshold weights and the threshold-specific predictive extreme value inferences.
The change from the MDI(0.6) prior to the Cauchy prior has changed little the threshold weights.  
The main difference is that the highest thresholds benefit more greatly from the increased prior information than the lower thresholds, resulting in a slight increase in their estimated threshold weights.
For the highest thresholds the Cauchy prior has prevented the very unrealistic estimates that were obtained under the MDI(0.6) prior.
This can also be seen by comparing the plots on the right of Figures \ref{fig:BMA} and \ref{fig:Cthresh}: 
the grey curves corresponding to high thresholds have shifted to the left, that is, towards giving higher density to smaller values of $M_N$, with a similar knock-on effect on the threshold-averaged black curves.

\section{Discussion}
We have proposed new methodology for extreme value threshold selection based on a GP model for threshold excesses.
It can be used either to inform the choice of a `best' single threshold or to reduce sensitivity to a particular choice of threshold by averaging extremal inferences from several thresholds, weighting thresholds with better cross-validatory predictive performance more heavily than those with poorer performance.
The simulation study in section \ref{sec:sim2} shows that the estimated threshold weights behave as expected in cases where the GP model holds exactly above some threshold and illustrates the potential benefit of averaging different estimated tail behaviours to perform extreme value extrapolation.

The methodology has been applied to significant wave height datasets from the northern North Sea and the Gulf of Mexico.
For the latter dataset the highest thresholds result in physically unrealistic extrapolation to long future time horizons.
Averaging inferences over different thresholds avoids basing inferences solely on one of these thresholds,
but we also explored how the incorporation of basic prior information can be used to address this problem.
Stronger prior information about GP model parameters, or indeed prior information about the threshold level itself, could also be used.

In common with many existing methods we need a set of thresholds $u_1, \ldots, u_k$ or an interval $(u_{min}, u_{max})$ for threshold,
where $u_k$ (or $u_{max}$) does not exceed the highest threshold from which it is judged that meaningful inferences can be made.
This is also true of other methods, such as those that assess formally stability of parameter estimates and those for which the form of a standard extreme value model is extended to model data below the threshold.
In the latter case results can be sensitive to the form of the model specified below the threshold and to $u_{min}$.

The fact that our methodology is based on inferences from standard unmodified extreme value models makes it relatively amenable to generalization. 
In significant wave height examples considered in this paper it is standard to extract event maxima from raw data, thereby producing observations that are treated as approximately independent.  Otherwise, data may exhibit short-term temporal dependence at extreme levels, leading to clusters of extremes.  
In on-going work we are extending our general approach to this situation and to deal with other important issues: the presence of covariate effects; the choice of measurement scale; and inference for multivariate extremes. 

\section*{Acknowledgments}
We thank Richard Chandler, Kevin Ewans, Tom Fearn and Steve Jewson for helpful comments.
Nicolas Attalides was funded by an Engineering and Physical Sciences Research Council studentship while carrying out this work.

\bibliography{NAJ2015}
\bibliographystyle{rss}

\end{document}